\documentclass[12pt,preprint]{aastex}
\usepackage{epsfig}
\usepackage{natbib}
\usepackage{graphicx}
\usepackage{slashbox}
\usepackage{multirow}
\usepackage{lscape}
\usepackage{mathrsfs,amssymb}
\usepackage{amsmath}
\newcommand	\beq	{\begin{equation}}	%{\begin{displaymath}}
\newcommand	\eeq	{\end{equation}}	%{\end{displaymath}}
\newcommand       \Angstrom     {\,{\rm \AA}}
\newcommand       \AU           {\,{\rm AU}}
\newcommand       \cm           {\,{\rm cm}}
\newcommand       \mm           {\,{\rm mm}}

\newcommand       \erg          {\,{\rm erg}}

\newcommand       \g            {\,{\rm g}}
\newcommand       \K            {\,{\rm K}}
\newcommand       \pc           {\,{\rm pc}}

\newcommand       \s            {\,{\rm s}}

\newcommand       \yr       {\,{\rm yr}}

\newcommand       \Myr      {\,{\rm Myr}}

\newcommand       \um           {\mu{\rm m}}
\newcommand       \mum          {\,{\rm \mu m}}

\newcommand       \Teff         {T_{\rm eff}}
\newcommand       \msun         {\,{M_\odot}}
\newcommand       \Lsun         {\,{L_\odot}}

\newcommand       \Lstar        {L_\star}

\newcommand       \simali       {\sim\,}

\newcommand       \Msil           {M_{\rm sil}}
\newcommand       \Mcarb          {M_{\rm carb}}
\newcommand       \rhosil        {\rho_{\rm sil}}

\def    \amin		{a_{\rm min}}
\def    \amax		{a_{\rm max}}

\newcommand{\etal}{\textrm{et al.\ }}
\newcommand{\ie}{\textrm{i.e., }}
\newcommand{\eg}{\textrm{e.g., }}

\newcommand{\Msun}      {\,{M_\odot}}
\newcommand{\Mearth}    {\,{M_\earth}}

\newcommand       \apahmin      {a^{\rm PAH}_{\rm min}}
\newcommand	  \rmin	        {r_{\rm min}}
\newcommand	  \rmax	        {r_{\rm max}}

\newcommand \thisdisk {HD\,169142}

\shorttitle{Dust and PAHs in the HD\,169142 Disk}
\shortauthors{Seok \& Li}
\title{
Dust and Polycyclic Aromatic Hydrocarbon
in the Pre-Transitional Disk around HD\,169142 
}
\author{Ji Yeon Seok and Aigen Li}
\affil{Department of Physics and Astronomy,
        University of Missouri,
        Columbia, MO 65211, USA;
        {\sf seokji@missouri.edu,
             lia@missouri.edu}}

\begin{document}

\begin{abstract}
The pre-transitional disk around the Herbig Ae star
HD 169142 shows a complex structure
%consisting of an inner cavity, a bright ring, an annular
%gap, and an outer disk 
of possible ongoing planet formation
in dust thermal emission from the near infrared (IR)
to millimeter wavelength range. Also, a distinct
set of broad emission
features at 3.3, 6.2, 7.7, 8.6, 11.3, and $12.7\mum$,
commonly attributed to polycyclic aromatic
hydrocarbons (PAHs), are detected prominently
in the \thisdisk~disk. 
We model the spectral energy distribution (SED)
as well as the PAH emission features of the
HD 169142 disk simultaneously
with porous dust and astronomical-PAHs
taking into account the spatially resolved disk
structure. Our porous dust model consisting
of three distinct components
that are primarily concentrated in the inner
ring, middle ring, and outer disk, respectively, 
provides an excellent fit to the entire SED,
%except a slight deficiency at 7 mm, 
and the PAH model %including PAH molecules
%with an ionization fraction of $\simali$$60\%$
closely reproduces the observed PAH features. 
The accretion of ice mantles onto porous 
dust aggregates
occurs between $\simali$$16\AU$ and $60\AU$,
which overlaps with the spatial extent ($\simali$$50\AU$)
of the observed PAH emission features.
%The PAH emission features are extended up to
%$\simali$$50\AU$, which overlaps with the region where
%ice mantles accrete onto porous dust aggregate
%(between $\simali$$16\AU$ and $60\AU$). 
%PAHs and dust in the \thisdisk~disk are
%subject to continuous removal via the stellar
%radiation pressure and Poynting--Robertson drag.
%In addition, PAHs are also subject to photodissociation.
%Therefore, these materials must be continuously
%replenished by mutual collisions
%among larger bodies such as planetesimals
%and asteroids and/or the outgassing of cometary bodies. 
Finally, we discuss the role of PAHs in
the formation of planets possibly taking place
in the \thisdisk~system. 

\end{abstract}


\section{Introduction}\label{sec:intro}

Transitional disks are a subgroup of protoplanetary
disks, which are primarily characterized by
gaps of large inner cavities of a few tens of
astronomical units (AU) (\eg see Strom \etal1989).
These cavities are considered
to be nearly devoid of solid materials,
and several mechanisms have been explored
for the disk clearing phenomena
such as photoevaporation
(\eg Alexander \etal2006),
accretion toward the central star
(\eg Perez-Becker \& Chiang 2011),
grain growth and coagulation onto large bodies
(\eg Dullemond \& Dominik 2005),
and large-scale disturbing by orbiting companions
(\eg Papaloizou \etal2007, 
Andrews \etal2011,
Zhu \etal2011).
Recently, it is found that some of the transitional disks
have shown significant infrared (IR) excess emission
at $\simali$2--6$\mum$ implying the possible
existence of a remaining inner disk inside the cavity
($r\la 1\AU$).
These are classified as ``pre-transitional'' disks
(\eg Espaillat \etal2007), which would be an earlier
evolutionary stage of transitional disks.
Since the presence of such an inner disk
is in favor of the large-scale
dynamical disturbing by companions
(\eg Andrews \etal2011),
pre-transitional disks are of great interest 
to look for ongoing planet formation and to
investigate the associated physical and chemical
conditions. %associated with it.

\thisdisk~(also known as SAO 186777)
was first classified as a Vega-like star
due to its IR excess based on 
the broadband photometry of
the {\it Infrared Astronomical Satellite}
($IRAS$; Walker \& Wolstencroft 1988). 
Later, it was suggested to be 
a pre-main sequence star
at an earlier evolutionary stage than
Vega-type stars (Sylvester \etal1997)
and has been considered as a Herbig Ae
star with a stellar mass of
$M_\star=1.65\msun$ (Blondel \& Djie 2006)
located at $\simali$$145 \pc$ away (\eg Sylvester \etal1997).
The \thisdisk~disk shows complex structures including
at least two cavities (or gaps), one ring structure, and
an outer disk, which have been spatially
resolved (\eg Honda \etal2012, Quanz \etal2013,
Osorio \etal2014, Momose \etal2015).
Also, since its spectral energy distribution (SED) shows
a significant near-IR excess at $2\mum\la\lambda\la6\mum$
indicative of the presence of an inner disk
(\eg Honda \etal2012,
Osorio \etal2014, Wagner \etal2015),
the \thisdisk~disk becomes one of the
exemplary pre-transitional disks.
Furthermore, a point-like feature within the inner cavity
has been detected by two independent
observations with the NACO camera
at the Very Large Telescope
(VLT; Biller \etal2014, Reggiani \etal2014),
%($\simali$$0\farcs11\pm0\farcs03$ or $\simali$$16\AU$
%at $d=145 \pc$, Biller \etal2014,
%$\simali$$0\farcs156\pm0\farcs032$ or 
%$\simali$$23\AU$ at $d=145 \pc$, Reggiani \etal2014),
which implies the possible occurrence of
(multiple) planet formation in the disk.

\thisdisk~was reported as the first star among Vega-like to have
``unidentified infrared'' (UIR) bands in its circumstellar disk
(Sylvester \etal1994) 
as revealed through the broad emission features at
7.7, 8.6, and 11.3 $\mum$ obtained with the 7.5--$13.5\mum$
Cooled Grating Spectrometer 3 (CGS3) 
at the United Kingdom Infrared Telescope (UKIRT). 
Since then a number of observational and/or theoretical
studies have examined
these features as well as other associated
ones such as the $3.3\mum$ feature in the \thisdisk~disk
(\eg Sylvester \etal1996, 1997, 
Meeus \etal2001,
Acke \& van der Ancker 2004,
Smith \etal2004,
Sloan \etal2005, 
Keller \etal2008, 
Maaskant \etal2014). 
%and the spatially extended 3.3 and $11.3\mum$
%features have been resolved (Habart \etal2006,
%Maaskant \etal2013).
Habart et al.\ (2006) obtained 
the $\simali$3.2--3.76$\mum$ $L$-band 
long-slit spectrum of the HD\,169142 disk 
with the NACO adaptive optics system
at the VLT and found that the 3.3$\mum$ UIR feature 
is spatially extended up to $\simali$50$\AU$. 
%... adaptive optics system NAOS-CONICA (NACO) at the VLT, 
% ...the spectral resolution $\sim$700. 
More recently, Maaskant et al.\ (2013) 
obtained the $\simali$8--13$\mum$ N-band 
long-slit spectrum of this disk 
with the VLT Spectrometer and Imager for the Mid--Infrared
(VISIR) also at the VLT
and found that the 11.3$\mum$ UIR feature 
is spatially extended as well.

The UIR features are
commonly attributed to polycyclic
aromatic hydrocarbon (PAH) molecules
(L\'{e}ger \& Puget 1984, Allamandola \etal1985),
which play an important role 
in the physical and chemical evolution of
protoplanetary disks. PAHs are excited by
ultraviolet (UV) and visible photons, 
of which the absorbed energy, in turn, 
is re-emitted through the vibrational
relaxation of PAHs via available internal vibrational
modes including the C--H stretching mode at
$3.3\mum$, C--C stretching at 6.2 and $7.7\mum$,
C--H in-plane bending at $8.6\mum$, and
C--H out-of-plane bending at $11.3\mum$
and $12.7\mum$. This makes the PAH emission
features an important diagnostic of the physical
conditions of the disk.
Photoelectrons of PAH molecules are the
dominant heating source of the gas in the
surface layers of the disk. Furthermore,
PAHs probe the presence of very small grains,
which can infer the dust processing in the disk
such as settling and coagulation, and these
processes are essential to form planetesiamls
as a building block of planet formation.    

In the context of the recent detection of a planet candidate
as well as a rich set of the prominent PAH features, 
the \thisdisk~disk provides an excellent laboratory
to investigate the physical and chemical properties
of PAHs and dust associated with the formation of
planets. We perform an extensive modeling
of the PAH features adopting the
astro-PAH model of Li \& Draine (2001) and
Draine \& Li (2007), which takes
precise PAH chemistry into account.
In the meantime, 
we model the dust continuum simultaneously using
the porous dust model of Li \& Lunine (2003a, 2003b)
and also infer the disk geometry. 
In the following sections, we briefly describe
the observational data in Section \ref{sec:data}
and explain the adopted models for dust and PAHs
in Section \ref{sec:model}
taking the previously-known disk geometry into
account.
We present the model results in Section \ref{sec:result}
and discuss the physical properties of dust and PAHs
in the \thisdisk~disk in Section \ref{sec:discus}.
Finally, we summarize the major results 
in Section \ref{sec:sum}.

\section{Data}\label{sec:data}

%\thisdisk~has been observed
%at wide wavelengths by various
Multi-wavelength photometric data and 
IR spectra of the \thisdisk~disk
are collected from the literature
to model the SED
and the PAH features simultaneously.
We adopt the UV fluxes from the
{\it International Ultraviolet Explorer} ($IUE$) archive,
the {\it UBVRI} optical fluxes from Sylvester \etal(1996),
the {\it JH$K_S$} near-IR photometry from the 2MASS
{\it All-sky Point Source Catalog},\footnote{%
  http://www.ipac.caltech.edu/2mass/releases/allsky
  }
mid- to far-IR fluxes 
obtained with ground-based telescopes
including
UKIRT (Sylvester \etal1996), 
Keck II (Jayawardhana \etal2001), 
Gemini (Mari{\~n}as \etal2011), and 
Subaru (Honda \etal2012)
as well as space telescopes
including 
the {\it Wide-field Infrared Survey Explorer} 
($WISE$),\footnote{%
  http://wise2.ipac.caltech.edu/docs/release/allsky
  }
{\it AKARI},\footnote{%
  http://www.ir.isas.jaxa.jp/AKARI/Observation/PSC/Public
  } 
{\it IRAS}, and
the {\it Herschel Space Telescope} (Meeus \etal2010),
sub-millimeter (submm) fluxes obtained with 
the Submillimetre Common-User Bolometer Array
(SCUBA) on the James Clerk Maxwell Telescope
(JCMT) from Sandell \etal(2011),
and millimeter (mm) data obtained with UKT14
on JCMT (Sylvester \etal1996) and with 
the Very Large Array (VLA; Osorio \etal2014).
All the photometric fluxes and uncertainties
are listed in Table \ref{tab:flx} with their
references.
For IR spectroscopy, we take
the $\simali$2.4--15$\mum$ spectrum 
of Acke \& van den Ancker (2004)
obtained with 
the Shorter Wavelength Spectrometer (SWS)
on board the {\it Infrared Space Observatory} (ISO)
and the $\simali$5--36$\mum$ spectrum
of Keller \etal(2008)
obtained with the Infrared Spectrograph (IRS)
on board the {\it Spitzer Space Telescope}.

\begin{figure*}[htbp]
\epsscale{1.00}
\plotone{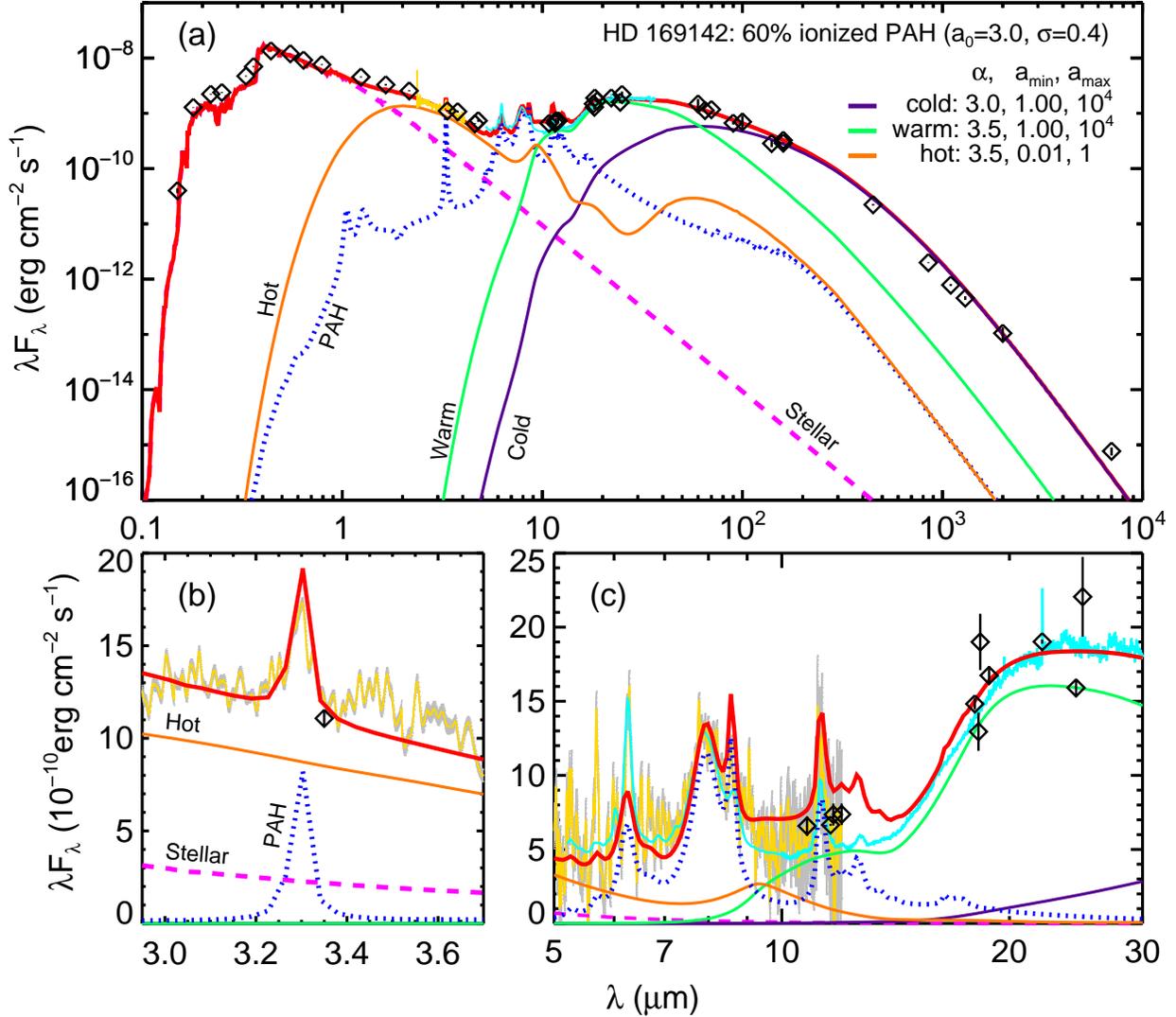}
\caption{\footnotesize
         \label{fig:sed}
         (a) Spectral energy distribution of HD\,169142. 
          Photometric fluxes are marked with diamonds,
          and uncertainties are overlaid with bars
          (see Table \ref{tab:flx}).
          \textit{ISO}/SWS spectrum
          (Acke \& van der Ancker 2004)
          and \textit{Spitzer}/IRS spectrum
          (Keller \etal2008) are shown as yellow and
          cyan lines, respectively.
          Our best-fit model spectrum is depicted with
          a red solid line together with individual components:
          stellar emission (magenta dashed line),
          PAH features (blue dotted line), and
          three dust components
          (hot, warm, and cold dust marked by 
          orange, green, and purple lines, respectively). 
          See text for details.         
         (b)--(c) Same as panel (a), but closeups of 
         the \textit{ISO}/SWS spectrum 
         and the \textit{Spitzer}/IRS spectrum, respectively.
	Uncertainties of the {\it ISO} spectra are superposed in gray.
	Note that the \textit{ISO}/SWS spectrum 
	at wavelengths longer than $5\mum$ shown
	in panel (c) is
	not used for modeling but is overlaid here
	to demonstrate the consistency between the 
	\textit{ISO}/SWS and \textit{Spitzer}/IRS
	spectrum.
         }
\end{figure*}

The observed SED at the entire wavelengths
together with the IR spectra
is shown in Figure \ref{fig:sed}(a).
The stellar atmospheric model spectrum
is overlaid for the case of $\Teff=8250$ K 
and log $g=4.0$ (Kurucz 1979).
We adopt an interstellar extinction of $A_V=0.01$
(Blondel \& Djie 2006)
toward \thisdisk~located at the distance of 
$d\approx145$ pc (\eg Sylvester \etal1997).
All stellar parameters adopted in this paper
are summarized in Table \ref{tab:para}.

The PAH emission features detected 
in the {\it ISO}/SWS and {\it Spitzer}/IRS
spectra are illustrated 
in Figures \ref{fig:sed}(b), (c). 
The major PAH features
at 3.3, 6.2, 7.7, 8.6, and $11.3\mum$ 
are prominent, and several minor features
such as the 3.43, 6.87, and 7.23$\mum$ features
attributed to aliphatic hydrocarbon
are also detected.
No silicate feature is found 
in the {\it Spitzer}/IRS spectrum.
As noted by Sloan \etal(2005),
the 6.2 and $7.7\mum$ features are shifted to
6.3 and 7.9$\mum$, respectively,\footnote{%
  The shift from $7.7\mum$ still remains
  an enigma. Sloan \etal(2005) suggest that the
  $7.7\mum$ shift is not sensitive to the charge
  of PAHs and speculate the influence of the PAH size.
  Later, a correlation between
  the peak wavelength of the $7.7\mum$ feature
  and the effective temperature of the exciting star
  is found (Sloan \etal2007): 
  as the exciting star is cooler, the $7.7\mum$
  feature is shifted to longer wavelengths.
  In addition, it is suggested that the shift may be
  associated with the presence of circumstellar material
  (Tielens 2008).
}
which are
common for the so-called ``class B'' PAH spectra
(Peeters \etal2002). Class B sources are
typically planetary nebulae, 
post-asymptotic giant branch stars, 
and isolated Herbig Ae/Be stars 
and are likely to be dominated
by relatively pure PAHs (Peeters \etal2002).

Recently, the variability of the disk emission of \thisdisk~
has been reported by Wagner \etal(2015).
While the variability mostly occurs 
at the near-IR wavelengths
($\simali$$1.5$--$10\mum$, 
up to 45\% over $\simali$$10\yr$),
it is found that the intensities of 
the PAH features do not change with
respect to their underlying continuum.
As we aim to focus on the PAH emission
and the variability-associated activity 
takes place within the $r\lesssim1\AU$ region 
of the disk (Wagner \etal2015),
it is unlikely that our results would be affected.
Thus, we have shifted the first segment
(2.36--$4.08\mum$) of the {\it ISO}/SWS spectrum
to fainter absolute fluxes by $\simali$$0.44$ Jy
to be consistent with the {\it WISE} photometric fluxes
(see Figures \ref{fig:sed}(b)).
The {\it ISO}/SWS spectrum at longer wavelengths
(\ie $\lambda\ga5\mum$) where the {\it Spitzer}/IRS
spectrum overlaps is not used for modeling but is
overlaid in Figure \ref{fig:sed}(c) to illustrate
its agreement with the {\it Spitzer}/IRS spectrum.

%%%%%%%% Table 1: SED %%%%%%%%

\begin{deluxetable}{ccccc}
\center
\tablecaption{\label{tab:flx}
	Photometric Data of the HD\,169142 System}
\tablehead{ \colhead{Wavelength} & \colhead{Flux} 
& \colhead{Uncertainty} &\colhead{Telescope/Filter} 
& \colhead{Reference} \\ 
 \colhead{(\micron)} & \colhead{(Jy)} 
 & \colhead{(Jy)} & \colhead{} & \colhead{}} 
\startdata
%#Initially taken from Table 2A from Maaskant et al.2013 A\&A 555, A64 \\
 0.15 & 0.002 & 0.000 & {\it IUE} & 1 \\
 0.18 & 0.077 & 0.001 & {\it IUE} & 1 \\
 0.22 & 0.163 & 0.003 & {\it IUE} & 1 \\
 0.25 & 0.199 & 0.004 & {\it IUE} & 1 \\
 0.33 & 0.519 & 0.010 & {\it IUE} & 1 \\
 0.36 ($U$) & 0.841 & 0.019 & \dots & 2 \\
 0.44 ($B$) & 1.986 & 0.026 & \dots & 2 \\
 0.55 ($V$) & 2.207 & 0.020 & \dots & 2 \\
 0.64 ($R$) & 1.963 & 0.026 & \dots & 2 \\
 0.79 ($I$) & 2.004 & 0.041 & \dots & 2 \\
 1.24 ($J$)& 1.880 & 0.040 & 2MASS & 3 \\
 1.65 ($H$)& 1.810 & 0.060 & 2MASS & 3 \\
 2.16 ($Ks$) & 1.840 & 0.040 & 2MASS & 3 \\
 3.35 & 1.238 & 0.047 & {\it WISE} & 4 \\
 3.77 ($L$) & 1.349 & 0.064 & UKIRT & 2 \\
 4.60 & 0.995 & 0.021 & {\it WISE} & 4 \\
 4.78 ($M$) & 1.172 & 0.055 & UKIRT & 2 \\
 10.8 ($N$) & 2.370 & 0.237 & Keck II & 5 \\
 11.6 & 2.575 & 0.028 & {\it WISE} & 4 \\
 11.7 & 2.870 & 0.290 & Gemini & 6 \\
 12.0 & 2.950 & 0.290 & {\it IRAS} & 7 \\
 18.0 & 8.900 & 0.230 & {\it AKARI}/IRC & 8 \\
 18.2 & 7.860 & 0.786 & Keck II & 5 \\
 18.3 & 11.60 & 1.16 & Gemini & 6 \\
 18.8 & 10.50 & 0.40 & Subaru & 9 \\
 22.1 & 14.02 & 0.10 & {\it WISE} & 4 \\
 24.5 & 13.00 & 0.50 & Subaru & 9 \\
 25.0 & 18.39 & 2.24 & {\it IRAS} & 7 \\
 60.0 & 29.57 & 6.81 & {\it IRAS} & 7 \\
 65.0 & 24.45 & 0.10 & {\it AKARI}/FIS & 10 \\
 70.0 & 27.35 & 0.03 & {\it Herschel} & 11 \\
 90.0 & 19.99 & 1.23 & {\it AKARI}/FIS & 10 \\
 100 & 23.37 & 6.74 & {\it IRAS} & 7 \\
 140 & 13.32 & 1.89 & {\it AKARI}/FIS & 10 \\
160 & 15.47 & 0.46 & {\it AKARI}/FIS & 10 \\
160 & 17.39 & 0.05 & {\it Herschel} & 11 \\
 450 & 3.340 & 0.115 & JCMT/SCUBA & 12 \\
 850 & 0.565 & 0.010 & JCMT/SCUBA & 12 \\
1100 & 0.287 & 0.013 & JCMT/UKT14 & 2 \\
1300 & 0.197 & 0.015 & JCMT/UKT14 & 2 \\
2000 & 0.070 & 0.019 & JCMT/UKT14 & 2 \\
7000 & 0.0018 & 0.0003 & VLA & 13 \\
\enddata
%\tablecomments{}
\tablerefs{(1) {\it IUE} archival data; 
(2) Sylvester \etal(1996);
(3) 2MASS {\it All-sky Point Source Catalog};
(4) {\it WISE All-sky Data Release Catalog};
(5) Jayawardhana \etal(2001);
(6) Mari{\~n}as \etal(2011);
(7) {\it IRAS Point Source Catalog};
(8) {\it AKARI/IRC All-sky Survey Point Source Catalog} 
	(Version 1.0, Ishihara et al. 2010);
(9) Honda \etal(2012);
(10) {\it AKARI/FIS All-Sky Survey Bright Source Catalog}
	(Version 1.0);
(11) Meeus \etal(2010);
(12) Sandell \etal(2011);
(13) Osorio \etal(2014)} 

\end{deluxetable}
%%%%%%%% Table 1: SED %%%%%%%%

%%%%%%%%%%%%%%% Figure 1: SED %%%%%%%%%%%%%%%
%%%%%%%%%%%%%%% Figure 1: SED %%%%%%%%%%%%%%%

%%%%%%%% Table 2: Stellar Parameters %%%%%%%%
\begin{deluxetable}{cccc}
\center
\tablecaption{\footnotesize
	\label{tab:para}
	Stellar Parameters of the HD 169142 System}
\tablehead{\colhead{Parameter} &\colhead{Unit} & \colhead{Value} &  \colhead{Reference}  }

\startdata
Sp. Type & & {\bf A5Ve}/A7Vz/A9III/IVe & 1, 2, 3 \\
log $g$ & & 4.2/$3.7\pm0.1$/{\bf 4.0-4.1} & 1, 3 ,4 \\
$T_{\rm eff}$ & K & 8400/$7500\pm200$/{\bf 8250} & 1, 3, 5\\
$M_\star$ & $\Msun$ & {\bf 1.65} & 2 \\
$R_\star$ & $R_\sun$ & 1.59/{\bf 1.6}/1.9 & 2, 4, 5  \\
$L_\star$ & $L_\sun$ & {\bf 8.55}/14.5/$9.4\pm5.6$ & 2, 6, 7 \\
$M_{\rm gas}$ & $10^{-3}\Msun$ & {\bf 6.0-30}/$5.0\pm2.0$& 8, 9 \\
Age ($\tau_\star$) & Myr & $\simali$$12$/$7.7\pm2.0$/${\bf 6^{+6}_{-3}}$ & 2, 7, 10\\
$A_V$ & & {\bf 0.01}/0.43/0.00 & 2, 6, 7 \\
Distance & pc & 151/{\bf 145}/$145\pm43$ & 2, 5, 7 \\ 
\enddata

\tablecomments{Adopted values in this work are in bold.} 
\tablerefs{(1) Dunkin et al.\ (1997); 
(2) Blondel \& Djie (2006); 
(3) Guimar{\~a}es et al.\ (2006); 
(4) Meeus et al.\ (2010); 
(5) Sylvester et al.\ (1997); 
(6) Acke \& van den Ancker (2004); 
(7) Meeus et al.\ (2012);
(8) Pani{\'c} et al.\ (2008);
(9) Meeus et al.\ (2010);
(10) Grady et al.\ (2007). 
}

\end{deluxetable}
%%%%%%%% Table 2: Stellar Parameters %%%%%%%%%%%%

\section{Model}\label{sec:model}

We adopt the porous dust model of Li \& Lunine (2003a, 2003b)
and the astro-PAH model of Li \& Draine (2001)
and Draine \& Li (2007) for modeling
the dust continuum as well as the PAH features
present in the observed IR spectra and SED. 
The modeling of dust and PAH emission
for \thisdisk~is similar to the modeling done
on the HD 34700 debris disk in Seok \& Li (2015),
to which the reader is referred for more details.
In addition, since the dust thermal IR emission
in the \thisdisk~disk has been spatially
resolved previously, we elaborate
the spatial (radial) distribution of dust in the disk.
We briefly describe the adopted models 
with special attention paid to the effect of 
the disk geometry.

%%%%%%%%%%%%% Figure 2: Disk Sketch %%%%%%%%%%%%%%
\begin{figure}[tbp]
\epsscale{1.00}
\plotone{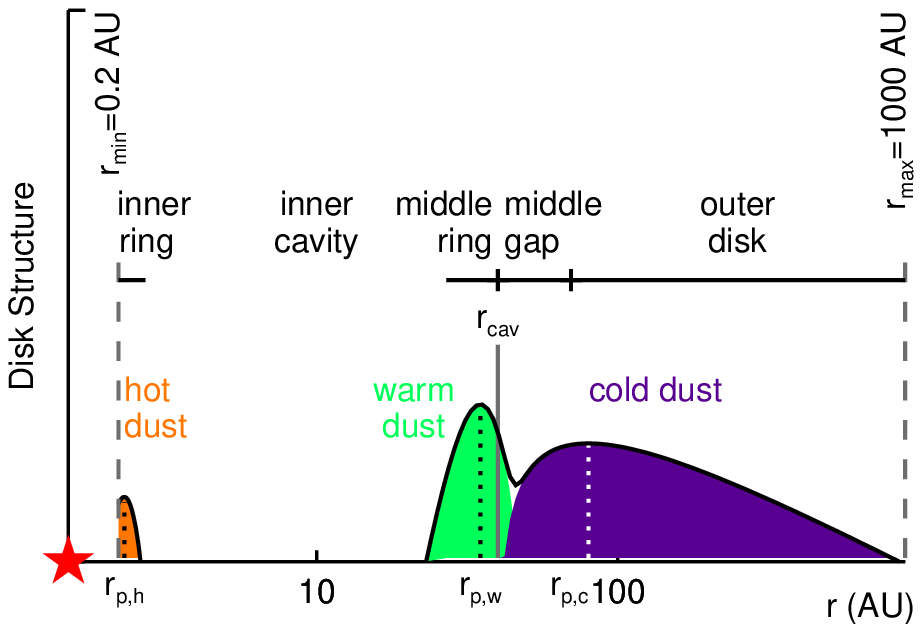}
\caption{\footnotesize
         \label{fig:disk}
	Schematic diagram of the \thisdisk~disk structure
	adopted in this work.
        The disk consists of an inner ring, a middle ring, 
        and an outer disk, which are separated by
        an inner cavity and a middle gap.
        Three dust components, hot, warm, and
        cold dust shown in Figure \ref{fig:sed}, are supposed
        to originate mainly from the inner ring, the middle ring, and
        the outer disk, respectively. Spatial distributions
        of the dust components in these structures are
        approximated by analytic formulae 
        (Equations (\ref{eq:dndr_h})--(\ref{eq:dndr_c})) 
        which peak at $r_{\rm p, h}$, $r_{\rm p, w}$,
        and $r_{\rm p, c}$, respectively. 
        The disk structures show in the diagram reflect
        the analytic formulae in an arbitrary vertical scale.
        The inner boundary, $\rmin=0.2\AU$, is taken to
        be the sublimation radius for dust
        (i.e., inside this radius the disk is dust-free), 
        and the outer boundary, $\rmax$, is assumed
        to be $1000\AU$. The inner boundary of the middle gap,
        $r_{\rm cav}$ is taken to be $40\AU$
        (Osorio \etal2014).
        }
\end{figure}         

%%%%%%%%%%%%% Figure 2: Disk Sketch %%%%%%%%%%%%%%

\subsection{Disk Geometry}\label{sec:disk}

Various observations that spatially resolved
the \thisdisk~disk 
(\eg Quanz \etal2013, 
Osorio \etal2014, 
Momose \etal2015)
enable us to better constrain the
spatial distribution of dust in the disk.
These observations have revealed
a complex structure of the disk by direct imaging: 	
an inner cavity ($1\AU \la r\la25\AU$), 
a middle ring ($25\AU \la r \la 40\AU$), 
a middle gap ($40\AU \la r \la 70\AU$), 
and an outer disk ($r\ga75\AU$).
Also, an inner ring (or inner disk, $r\la1\AU$)
is inferred
by the $\simali$2--6$\mum$ near-IR 
excess emission continuum.
Based on these observational results,
the disk structure that we adopt is schematized
in Figure \ref{fig:disk}.
With this geometry, we suppose that the
dust in the \thisdisk~disk consists of 
three distinct components:
hot, warm, and cold components.
Each component is mainly located in one part
of the disk:
hot dust originates from the inner ring
very close to the central star,
warm dust peaks at the middle ring,
and cold dust is mainly distributed throughout
the outer disk beyond the middle gap.
As shown in Figure \ref{fig:disk} as well as
in previous observations (\eg Grady \etal2007,
Quanz \etal2013, Osorio \etal2014,
Momose \etal2015), the middle gap is not really
a gap but a relatively depleted region.  
To formulate the spatial distribution of each component,
we adopt two Gaussian profiles for the hot and warm
components and one modified power-law for the cold
component. The latter is more physical than
a simple power-law and has been shown to be successful 
in reproducing the SED of the HD 34700 debris disk 
(see Equation (2) in Seok \& Li 2015).  
The spatial distribution of each component is defined as 
\beq
\label{eq:dndr_h}
\left(\frac{dn}{dr}\right)_{\rm h}\propto {\rm exp}
\left[-4\ {\rm ln}\ 2 \left(\frac{r-r_{\rm p, h}}{w_{\rm h}} \right)^2
 \right]~,~~\rmin\leq r \leq \rmax,
\eeq

\beq
\label{eq:dndr_w}
\left(\frac{dn}{dr}\right)_{\rm w}\propto {\rm exp}
\left[-4\ {\rm ln}\ 2 \left(\frac{r-r_{\rm p, w}}{w_{\rm w}} \right)^2
 \right]~,~~\rmin\leq r \leq \rmax,
\eeq

\begin{eqnarray}
\label{eq:dndr_c}
\left(\frac{dn}{dr}\right)_{\rm c}\propto\left\{ \begin{array}{l}
    0, \hspace{35mm}\rmin\leq r \leq r_{\rm cav}, \\ \\
   \left(1-\frac{r_{\rm cav}}{r}\right)^\beta
   \left(\frac{r_{\rm cav}}{r}\right)^\gamma~, ~~ r_{\rm cav}\leq r \leq \rmax, 
   \end{array} \right.
\end{eqnarray}
where $r_{\rm p,h}$ and $r_{\rm p, w}$ are, respectively, the
peak radial distances of the hot and warm dust components,
$w_{\rm h}$ and $w_{\rm w}$ are, respectively,
the width of the inner and
middle rings, and $r_{\rm cav}$ is the inner boundary of
the middle gap.
The cold dust component peaks
at $r_{\rm p, c}=r_{\rm cav}(\beta + \gamma)/\gamma$.
The inner boundary of the entire dust disk ($\rmin$) is defined
as where dust grains sublimate 
(\ie $T_{\rm dust}\ga1500\K$), which is 
$\rmin=0.2\AU$ in the \thisdisk~disk.
For the outer boundary, $\rmax=1000\AU$ is assumed
as our model is barely affected by $\rmax$
(see Li \& Lunine 2003b, Seok \& Li 2015),
but it is within a reasonable range considering
the distance to its companion (Grady \etal2007)\footnote{%
A weak-line T Tauri binary, 2MASS 18242929--2946559
is located $9\farcs3$ to the southwest of \thisdisk, which
corresponds to a projected separation of $\approx1160\AU$
at $d=145 \pc$ (Grady \etal2007).}. 
Then, the total spatial distribution is expressed as 
\beq
\label{eq:dndr}
\frac{dn}{dr}=\left(\frac{dn}{dr}\right)_{\rm h}+
\left(\frac{dn}{dr}\right)_{\rm w}+\left(\frac{dn}{dr}\right)_{\rm c}.
\eeq

Since previous studies have revealed the detail
structure of the \thisdisk~disk 
(\eg Honda \etal2012, Quanz \etal2013, 
Osorio \etal2014, Momose \etal2015),
several parameters of the spatial distributions
can be adopted accordingly.
For the cold dust, $r_{\rm cav}=40\AU$ is set
based on Osorio \etal(2014), and 
$\gamma=1.0$ is adopted based on previous
observations of scattered light in the near-IR.\footnote{%
  The radial profiles of 
  the scattered light at 1.1$\mum$ obtained with 
  the Near Infrared Camera and Multi-Object Spectrometer 
  (NICMOS) coronagraphy 
  on board the {\it Hubble Space Telescope}
  (\eg Grady \etal2007) 
  and the polarized intensity of 
  the {\it H}-band scattered light
  obtained with Subaru
  (\eg Momose \etal2015)
  have an $r^{-3}$-dependence. 
  As the stellar radiation field $\propto r^{-2}$,
  the remaining $r^{-1}$ is attributed to the radial
  dependence on the grazing angle between
  the scattering surface and the stellar light
  (see Appendix 2 of Momose \etal2015).
  Although the column density of the uppermost layer 
  of the dust disk where the scattered light is emitted 
  is related to the total column density indirectly,
  we adopt a surface density proportional to $r^{-1}$ here
  and fully examine the effect of varying $\gamma$
  in discussion (see Section \ref{sec:robst}).
%  this indicates a surface density 
 % of the emitters proportional to $r^{-1}$.
  } 
For the warm component, we vary $w_{\rm w}$
and $r_{\rm p, w}$ within a narrow range to be
consistent with the middle ring seen in the direct
imaging (\ie the middle ring at $25 \AU\la r \la 40 \AU$,
Quanz \etal2013, Osorio \etal2014).
Finally, there are five free parameters to
be constrained: $r_{\rm p, h}$, $w_{\rm h}$,
$r_{\rm p, w}$, $w_{\rm w}$, and $r_{\rm p, c}$.

%%%%%%%% Figure 3 %%%%%%%%%%%%
\begin{figure}[tbp]
\epsscale{0.50}
\plotone{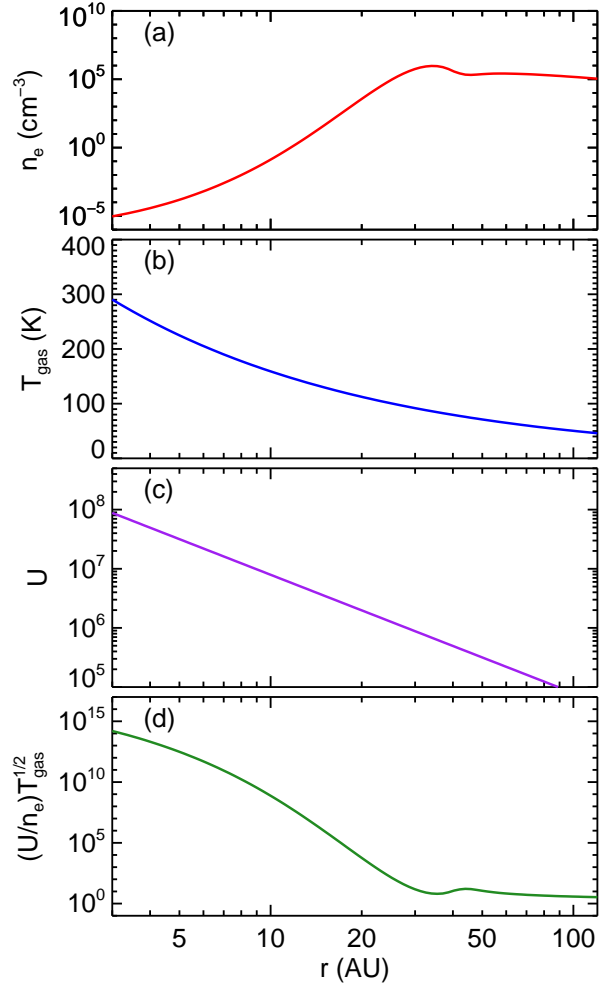}
\caption{\footnotesize
         \label{fig:ne} Radial profiles of (a) electron density,
         $n_{\rm e}$ ($\cm^{-3}$),
         (b) gas temperature, $T_{\rm gas}$ (K), 
         (c) starlight intensity, $U$ in unit of the 912$\Angstrom$--1$\mum$
             MMP83 ISRF (e.g., see Equation~(1) of
             Li \& Lunine 2003b), and
         (d) PAH ionization parameter, $UT^{1/2}_{\rm gas}/n_{\rm e}$ 
         (Bake \& Tielens 1994, Weingartner \& Draine 2001)
         as a function of distance, $r$ (AU),
         from the central star in the \thisdisk~disk. 
          }
\end{figure}
%%%%%%%% Figure 3: %%%%%%%%%%%%

\subsection{PAHs}\label{sec:pah}

For PAHs in the \thisdisk~disk, we adopt a log-normal
size distribution, $dn_{\rm PAH}/da$, characterized by
two parameters, the peak size ($a_0$)
and the width ($\sigma$) of the log-normal function
(Li \& Draine 2001, Li \& Lunine 2003b, and 
see also Equation (1) in Seok \& Li 2015).
The PAH size distribution ranges from
$\apahmin=3.5$ \AA, which is the minimum
size of PAHs of which can survive 
in the diffuse interstellar medium 
(ISM; Li \& Draine 2001). 

We adopt the PAH
absorption cross sections from Draine \& Li (2007).
Neutral and charged PAHs are considered separately
while neither cations from anions nor
multiply charged PAHs from singly charged PAHs 
are distinguished.
The absorption cross sections of the PAH features
are expressed as a series of Drude profiles
(Li \& Draine 2001, Draine \& Li 2007).
As applied to HD 34700 (Seok \& Li 2015), 
the peak wavelengths
and/or widths of some of the Drude profiles are
slightly modified (but the integrated area of each
profile is kept unchanged) to fit the PAH features
observed in the \thisdisk~disk,
which show some differences from 
those typically observed in the ISM
(see Section \ref{sec:data}).

%-PAH ionization
The coexistence of the prominent PAH features
at 3.3 and $11.3\mum$
and those at 6.2 and $7.7\mum$
in the observed IR spectra is indicative of
a mixture of neutral and ionized PAHs in
the \thisdisk~disk. The ionization-recombination
balance of PAHs is controlled
by the PAH ionization parameter, 
$UT^{1/2}_{\rm gas}/n_{\rm e}$,
where $U$ is the UV starlight intensity,
$T_{\rm gas}$ is the gas temperature,
and $n_{\rm e}$ is the electron density.
This parameter originates from two competing mechanisms,
the photoionization and the electron recombination
(Bake \& Tielens 1994, Weingartner \& Draine 2001).
Since the ionization parameter varies with the distance
from the central star, the charge state of PAHs
is expected to differ along the disk. 

Following Li \& Lunine (2003b), we derive
$n_{\rm e}$, $T_{\rm gas}$, and $U$ as a function
of $r$, the distance from the star 
(see also Section 3.1 in Seok \& Li 2015).
It is assumed
that electrons primarily result from
the cosmic-ray ionization of H$_2$, of which
the density is sufficiently high in the disk plane
(Pani\'{c} \etal2008).
Then, $n_{\rm e}(r)$ is
calculated from 
$n_{\rm e}\approx\tau_\star \varsigma_{\rm CR}n_{\rm H_2}$, 
where $\varsigma_{\rm CR}$ is 
the cosmic ionization rate, 
$\tau_\star$ is the stellar age, and $n_{\rm H_2}(r)$ is
the $\rm H_2$ density as a function of $r$. 
We adopt
$\varsigma_{\rm CR}\approx3\times10^{-17}\s^{-1}$
and $\tau_\star\approx6 \Myr$ 
(Grady \etal2007, see also Table \ref{tab:para}).
To derive $n_{\rm H_2}(r)$, we first obtained
the total H$_2$ mass ($M_{\rm H_2}$) from 
the total gas mass of the \thisdisk~disk ($M_{\rm gas}$)
in the literature (\ie $M_{\rm gas}\approx6\times10^{-3}\Msun$
from Pani\'{c} \etal2008).\footnote{%
 In Pani\'{c} \etal(2008), $M_{\rm gas}$ is 
 measured based on CO observations, and
 the measurement depends on the abundance of
 $^{12}$CO with respect to H$_2$ inevitably.
 Pani\'{c} \etal(2008) adopted a range of
 conversion factors (2.0--9.5)$\times10^{-4}$
 reported in the literature. We adopt their results
 obtained with the minimum conversion factor
 (\ie $M_{\rm gas}=6.0\times10^{-3}\Msun$),
 which gives the most consistent value with
 what Meeus \etal(2010) derived
 based on the far-IR observations of {\it Herschel}
 (see Table \ref{tab:para}).
}
Assuming that the spatial distribution of H$_2$
follows that of dust with a constant dust-to-H$_2$
mass ratio
(see Equations (\ref{eq:dndr_h})--(\ref{eq:dndr}))
and that the disk is in vertical hydrostatic
equilibrium,
we derive $n_{\rm H_2}(r)$ from $M_{\rm H_2}$.
With $\varsigma_{\rm CR}$ and $\tau_\star$,
this leads to
the radial distribution of electron density $n_{\rm e}$ 
as shown in Figure \ref{fig:ne}(a)
(see Appendix B in Li \& Lunine 2003b for details).
For $T_{\rm gas}$ shown in Figure \ref{fig:ne}(b), we take
$T_{\rm gas}(r)=(R_\star/2r)^{1/2}\Teff
\approx 503\times(r/\AU)^{-1/2}\K$,
where $R_\star\approx1.6R_\odot$ is
the stellar radius
for \thisdisk~(Meeus \etal2010).

The starlight intensity $U(r)$ is defined
as the intensity of the stellar radiation 
between 912$\Angstrom$ and 1$\mum$
at a distance $r$ from the central star of 
\thisdisk~with respect to that of the
local interstellar radiation field (ISRF) of
Mathis et al.\ (1983; MMP83).
Following Equation (1) in Li \& Lunine (2003b),
we obtain $U(r)\approx(2.81\times10^4 \AU/r)^2$
as shown in Figure \ref{fig:ne}(c).
Combining $n_{\rm e}$, $T_{\rm gas}$,
and $U$, we calculate the ionization parameter,
$UT^{1/2}_{\rm gas}/n_{\rm e}$ as
a function of radial distance ($r$). As shown
in Figure \ref{fig:ne}(d), PAHs near
the central star are highly ionized
(\ie high $UT^{1/2}_{\rm gas}/n_{\rm e}$)
whereas those far from the star are
mostly neutral or negatively charged
(\ie low $UT^{1/2}_{\rm gas}/n_{\rm e}$). 
A dip at $\simali$$30$--$40\AU$ and a bump
at $\simali$$50\AU$ in $UT^{1/2}_{\rm gas}/n_{\rm e}$
result from the spatial distribution of $n_{\rm e}$, 
which reflects the dust spatial distribution
in the disk (see Section \ref{sec:disk}).

\begin{figure}[tbp]
\epsscale{0.50}
\plotone{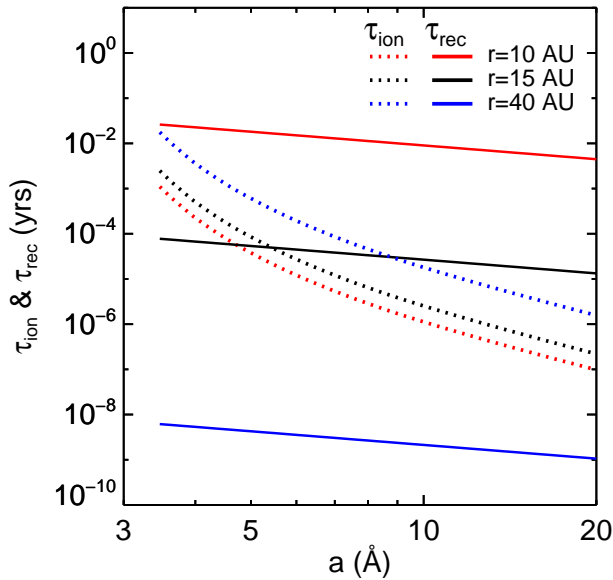}
\caption{\footnotesize
         \label{fig:ion}
         Timescales of photoionization ($\tau_{\rm ion}$, black lines)
         and electron recombination ($\tau_{\rm rec}$, red lines) as
         a function of PAH size, $a$ ($\Angstrom$). Timescales
         at given distances, $r=10$, 15, and 40 AU, are 
         represented by dashed, dotted, and solid lines, respectively.
          }
\end{figure}

To examine the charge state of PAHs
in the \thisdisk~disk further, 
we have calculated the photoionization rate
($k_{\rm ion}$) and the electronic recombination rate
($k_{\rm rec}$) at given distances taking into account
their dependences on PAH size
(see Equations (A7)-(A8) in Li \& Lunine 2003b).
Figure \ref{fig:ion} shows the timescales of photoionization
($\tau_{\rm ion}\equiv1/k_{\rm ion}$) and electron recombination
($\tau_{\rm rec}\equiv1/k_{\rm rec}$) as a function of PAH size
($a$ in \AA) at $r=10$, 15, and 40$\AU$.
While all PAH molecules at $\simali$$40\AU$ 
would experience electronic recombination immediately,
those at $\simali$$10\AU$ would be photo-ionized within
a shorter timescale against recombination.
Also, PAHs around $\simali$$40\AU$ may
become negatively-charged because abundant
free electrons are available to recombine with them.
The 3.3 and $11.3\mum$ PAH emission features
in the \thisdisk~disk were found to be spatially 
extended up to $\simali$$50\AU$
(Habart \etal2006, Maaskant \etal2013).
If PAHs are distributed from somewhere in the
inner gap to beyond (\eg $10\AU\la r \la50\AU$), 
and/or if some PAHs in the middle ring become
anion, we can naturally
explain that both neutral and charged PAHs exist
in the \thisdisk~disk as indicated by the observed
PAH features in the IR spectra.

To quantify the charge state of this mixture of
neutral and charged PAHs,
we examine the ionization fraction ($\phi_{\rm ion}$),
which is the probability of finding a PAH molecule in a 
nonzero charge state (Li \& Lunine 2003b).
The ionization fraction, $\phi_{\rm ion}$
depends on the PAH size ($a$), 
starlight intensity ($U$),
and electron density ($n_{\rm e}$).
In particular, a precise knowledge of $n_{\rm e}$ 
is needed for an accurate calculation of 
$\phi_{\rm ion}$ (Weingartner \& Draine 2001).
For simplicity, we first search for 
a constant $\phi_{\rm ion}$ for all PAH sizes
that can reproduce the observed PAH features closely.
Then, the best-fit value will be compared to
$\phi_{\rm ion}(a)$ as a function of PAH size ($a$)
derived with $n_{\rm e}(r)$ and
$U(r)$ shown in Figure \ref{fig:ne}
(see Section \ref{subsec:pah}).

Because of the single-photon heating nature
of PAHs, the spectral profiles of the PAH features
are independent of the starlight intensity $U$
(Draine \& Li 2001, Li \& Draine 2001).
Consequently, this indicates that the PAH abundance 
or the total mass of PAHs in the disk ($M_{\rm PAH}$)
becomes inversely proportional to $U$.

%%%%%%%% Figure 4 %%%%%%%%%%%%
%%%%%%%% Figure 4 %%%%%%%%%%%%

\subsection{Dust}\label{sec:dust}

Porous dust in the model of Li \& Lunine (2003a,b)
is assumed to be formed
via cold coagulation of pristine interstellar grains. 
Dust aggregates are highly fluffy (quantified by
porosity, $P$) and are composed of amorphous silicate
($\rhosil=3.5\g\cm^{-3}$, Draine \& Lee 1984) 
and carbonaceous materials ($\rho_{\rm carb}=1.8\g\cm^{-3}$,
Rouleau \& Martine 1991). Those at $T\la120\K$
regions also have mantles consisting
of H$_2$O-dominated ices
($\rho_{\rm ice}=1.2\g\cm^{-3}$, Li \& Greenberg 1998),
which have a slightly different porosity ($P^\prime$).
The mixing mass ratios are adopted to be
$\Mcarb/\Msil\approx0.7$ and
$M_{\rm ice}/(\Mcarb+\Msil)\approx0.8$,
where $\Mcarb$, $\Msil$, and $M_{\rm ice}$, 
are the total mass of the carbon, silicate, and ice
subgrains, respectively (Li \& Lunine 2003a).
It is found that the variation of porosity within 
the range of $0.8\la P \la 0.9$, which is expected
for dust aggregates via coagulation 
(Cameron \& Schneck 1965, Blum \& Wurm 2008)
does not affect the modeling significantly
(Seok \& Li 2015), so we adopt $P=0.90$
for the silicate-carbon aggregate dust
and $P^\prime\approx0.73$ for the ice-coated
silicate-carbon aggregate dust in \thisdisk~
(Li \& Lunine 2003a, 2003b).

Assuming that a dust grain is spherical in shape,
we adopt a power-law size distribution:
$dn/da\propto a^{-\alpha}~(\amin\le a \le \amax)$,
where $a$ is the radius of the spherical dust aggregate,
$\alpha$ is the power-law index, and $\amin$ and
$\amax$ are the lower and upper size cutoff.
Since the three dust components (hot, warm, and
cold dust in Section \ref{sec:disk}) are placed in
different physical conditions, in particular 
the hot dust at the inner ring, each could have
distinct size distributions.
We took $\amin=1\mum$ and $\amax=1\cm$
for the warm and cold dust components
(Li \& Lunine 2003b).
Varying the size cutoffs, in particular, $\amax$,
does not cause considerable changes in our results
as explored in Li \& Lunine (2003b)
and Seok \& Li (2015). However, 
for the warm component, $\amin=1\mum$
is preferred because smaller grains
produce a strong silicate feature at $10\mum$,
which is absent in the IR spectra of \thisdisk.
For the hot component, much smaller grains
($\amin=0.01 \mum$ and $\amax=1\mum$) are
adopted to account for the near-IR excess emission. 
Decreasing $\amin$ does not make much difference 
whereas larger $\amin$ enhances 
the $10\mum$ silicate feature. 
Hence, the power-law index of each component 
(\ie$\alpha_{\rm h}$, $\alpha_{\rm w}$, and $\alpha_{\rm c}$
for the hot, warm, and cold components, respectively)
is set to be a free parameter for the dust size distribution.
For a given set of these parameters combined with
those for the disk geometry (see Section \ref{sec:disk}), 
the total dust mass ($M_{\rm dust}$) is determined
by the flux level of the observed dust continuum
(see Equation (11) in Li \& Lunine 2003b).

%%%%%%%%%%%%%%%%% Results %%%%%%%%%%%%%%%%%
\section{Results}\label{sec:result}

%After modeling of the dust and PAH emission simultaneously,
Our best-fit model is shown in Figure \ref{fig:sed}, 
which provides an excellent fit to both the entire SED 
as well as the PAH emission features.
The best-fit model includes a mixture of neutral
and charged PAHs with an ionization fraction of
$\phi_{\rm ion}=0.6$. 
The relatively high $\phi_{\rm ion}$
indicates either that the PAH emission
from somewhere in the inner cavity contributes %($10\AU\la r\la25\AU$)
the total PAH intensity considerably 
or that a large fraction of PAHs in the middle ring are
negatively charged.
The former is because PAH cations are expected
to be mostly distributed at $\lesssim15\AU$
(see Section \ref{sec:pah} and Figure \ref{fig:ion}),
and this implies that the inner cavity devoid of small
dust grains might produce ionized PAH emission
contributing to its entire IR spectrum significantly 
even if their contribution to the total PAH mass
may not be significant. The latter is also
feasible since the PAH emission is known to 
be spatially extended up to $\simali$$50\AU$ 
(Habart et al.\ 2006, Maaskant et al.\ 2013)
and the electron recombination timescale 
in the middle ring is much shorter than 
the photoionization timescale
(see Figure \ref{fig:ion}).

\begin{figure}[tbp]
\epsscale{0.50}
\plotone{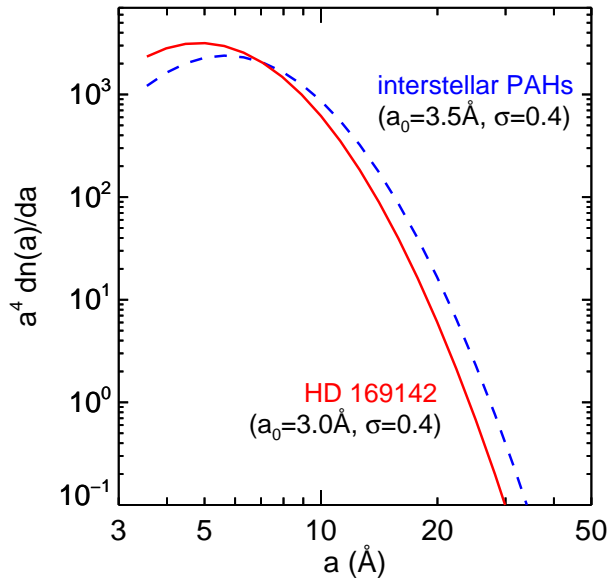}
\caption{\footnotesize
         \label{fig:pahsz}
         Log-normal size distribution for 
         the PAH population in the HD 169142 disk. 
         The best-fit model with $a_0\approx3.0\Angstrom$
         and $\sigma\approx0.4$ is compared with that of 
         the interstellar PAH model
         ($a_0\approx3.5\Angstrom$ 
          and $\sigma\approx0.4$, Li \& Draine 2001). 
          The size distribution is expressed by multiplying 
          $a^4$ to show the mass distribution per logarithmic 
          PAH radius.
          }
\end{figure}

The PAH size distribution is characterized by 
$a_0\approx3.0\Angstrom$ and $\sigma\approx0.4$
(Figure \ref{fig:pahsz}), which indicates that
a population of slightly smaller PAHs 
is preferred for the \thisdisk~disk 
compared with those in the diffuse ISM 
($a_0\approx3.5\Angstrom$
and $\sigma\approx0.4$, Li \& Draine 2001).

%%%%%%%%%%%%%%% Figure 5: dn/da for PAHs %%%%%%%%%%%%%%%
%%%%%%%%%%%%%%% Figure 5: dn/da for PAHs %%%%%%%%%%%%%%%

For the dust emission,
the best-fit model gives the dust size distribution
with power-law indices of $\alpha_{\rm h}\approx3.5$, 
$\alpha_{\rm w}\approx3.5$, and $\alpha_{\rm c}\approx3.0$
for the hot, warm, and cold components, respectively.
The flatter $\alpha_{\rm c}$ indicates that
larger grains are more abundant in
the cold dust component 
compared to the other components.

For the disk geometry, the hot dust component peaks
at $r_{\rm p, h}\approx0.3\AU$ with a width of 
$w_{\rm h}\approx0.1\AU$, the warm component peaks at
$r_{\rm p, w}\approx35\AU$ with $w_{\rm w}\approx10\AU$,
and the cold component peaks at $r_{\rm p, c}\approx80\AU$.
Though the spatial distribution of the hot component cannot
be verified due to the lack of spatially resolved 
observational data,
the other derived parameters 
are consistent with 
previous observations
(\eg a ring at $\simali$$14.5$--40$\AU$,
a gap at $\simali$$40$--70$\AU$, and an outer disk
at $\simali$$70$--250$\AU$ shown in Quanz \etal2013).
For the sake of convenience, we call this best-fit
model the standard model hereafter.

With the parameters of the standard model, 
we calculate a total dust mass of 
$M_{\rm dust}\approx176 \Mearth$
(i.e., $5.28\times10^{-4}\Msun$),
which is dominated by the cold component
($\simali$$99.48\%$). The hot and warm
components contribute only $\simali$0.06\% 
and $\simali$0.46\% of $M_{\rm dust}$, respectively.
Our estimate is in good agreement with
those reported in the literature
including $M_{\rm dust}\sim2.16\times10^{-4}\Msun$
(Pani{\'c} \etal2008) or
$\simali$$4\times10^{-4}\Msun$ (Honda \etal2012).
A total IR flux including both the dust and PAH emission
is derived as
$\int_{912\Angstrom}^{\infty} F_\lambda d\lambda
\approx5.4\times10^{-9}\erg\cm^{-2}\s^{-1}$, 
based on the standard model.
The hot, warm, and cold components contribute 
$\simali$$31\%$, 40\%, and 19\%, respectively,
while the PAH emission accounts for $\simali$$10\%$ of
the total IR emission. Unlike that the total dust mass
is dominated by the cold component, the contribution
of each component to the total IR emission
is comparable to each other. 
The total IR flux yields a total IR luminosity of
$L_{\rm IR}\approx3.5\Lsun$ at $d=145\pc$.
This gives a ratio of the IR luminosity to the
stellar luminosity of $L_{\rm IR}/\Lstar\approx0.41$
which is consistent with the estimates 
reported in previous studies
(\eg $L_{\rm IR}/\Lstar\approx0.42$, Meeus \etal2012).

%%%%%%%%%%%%%%% Figure 6: T(dust) %%%%%%%%%%%%%%%
\begin{figure}[tbp]
\epsscale{1.1}
\plottwo{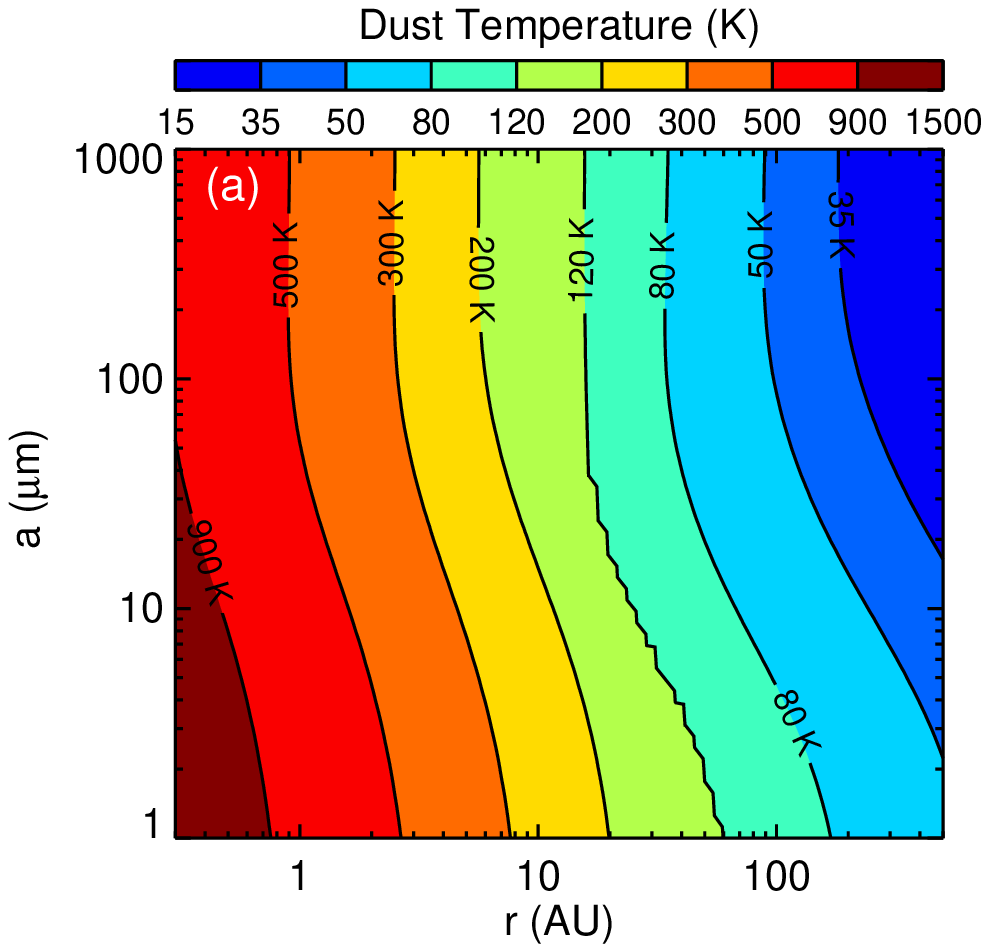}{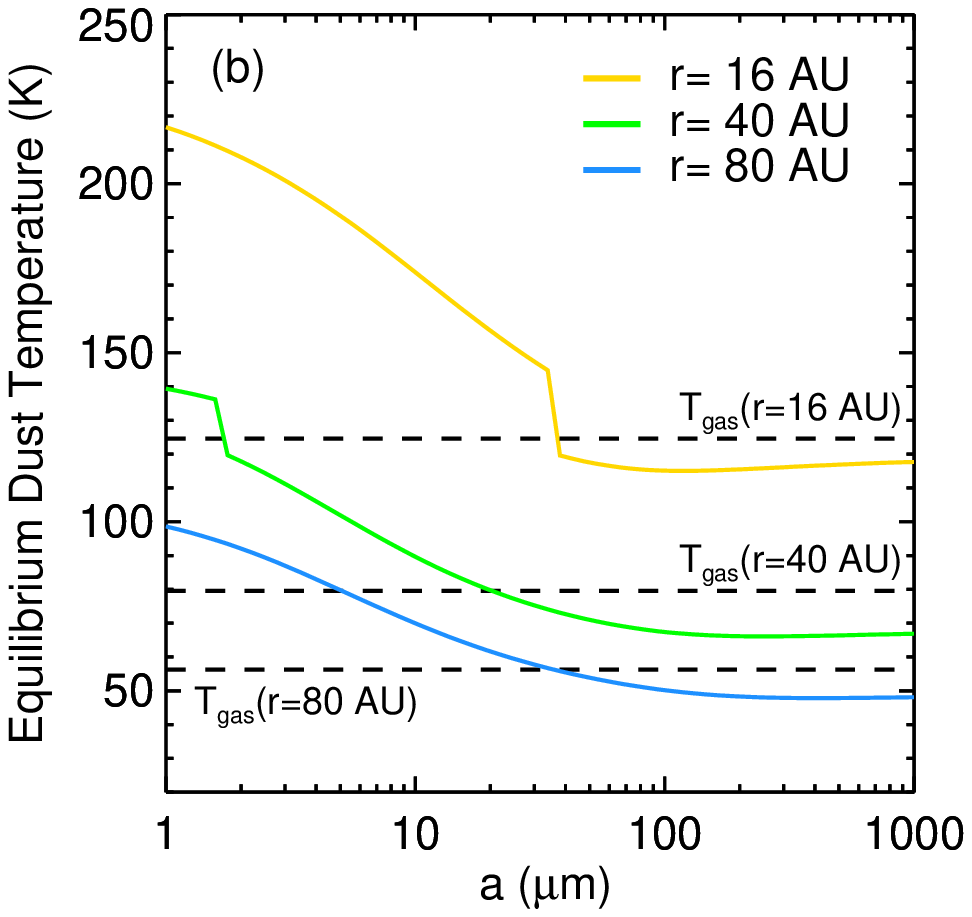}
\caption{\footnotesize
         \label{fig:Teq}
         (a) Variation of the equilibrium dust temperature $T(r, a)$
%         (of porosity $P=0.90$ for ice-free dust
%         and $P^{\prime}=0.73$ for ice-coated dust)  
         as a function of grain size ($a$ in $\um$) 
         and radial distance ($r$ in AU)
         from the central star in the \thisdisk~disk.
         Large grains ($\ga30\mum$) start to accrete
         icy mantles at $\simali$$16$ AU, and all grains
         become ice-coated beyond $\simali$$60\AU$. 
         (b) Dust temperature
         as a function of grain size 
         at given distances, $r=16\AU$ (yellow),
         $40\AU$ (green), and
         $80\AU$ (blue).    
         Dust temperatures at $r=16$ and 40\AU~show
         sudden jumps due to ice sublimation 
         when the temperature of an ice-coated grain 
         reaches $T\approx120\K$. 
	 For comparison, the gas temperatures 
         $T_{\rm gas}\approx503\left(r/{\rm AU}\right)^{-1/2}\K$ 
         at these distances are overlaid (dashed lines). 
         }
\end{figure}
%%%%%%%%%%%%%%% Figure 5: T(dust) %%%%%%%%%%%%%%%

Assuming that the dust grains are in thermal
equilibrium, we calculate the dust temperature ($T$)
as a function of grain size ($a$) and distance ($r$)
from the central star (Figure \ref{fig:Teq}(a)).
Porous dust with $P=0.9$ starts to have ice-coated subgrains
when its temperature becomes below $\approx$\,120$\K$.
Figure \ref{fig:Teq}(a) shows that
the ice sublimation of small grains 
starts to occur at $r\sim60\AU$ in the \thisdisk~disk,
and all icy mantles are expected to 
have sublimated at $\lesssim16\AU$.
Since the ice-coated porous dust ($P^{\prime}\approx0.73$)
is less absorptive due to the presence of the
more transparent ice mantles of subgrains,
the dust temperature varies abruptly
at the transition between icy and non-icy dust grains.

The transition can be seen in Figure \ref{fig:Teq}(b)
more clearly. It shows that the temperature of a dust
grain depends on its size and the distance from
the central star. The temperature
profiles at $r=16$ and $40\AU$ show rapid
changes around $T\approx120\K$. When a dust grain
in a disk accretes volatile molecules from the gas-phase
to form an ice mantle, its location (\ie the radial
distance from the central star) is defined as the
``snowline''. The snowline plays an important role
in the spatial distribution of (free-flying) PAHs
and the PAH chemistry as PAH molecules might
easily stick on the ice mantle of dust grains
and then will not emit the observable emission features.
As shown in Figure \ref{fig:Teq}(a) and (b), the
exact snowline ranges from $\simali$$16$ to $60\AU$
depending on the grain size. 
Recalling that the 3.3 and 11.3$\mum$ 
PAH emission features are detected 
in the disk with a spatial extent of
$\simali$$50\AU$ 
(Habart \etal2006, Maaskant et al.\ 2013),
the spatial distribution of PAH molecules 
in the \thisdisk~disk seems to closely correlate
with the snowline of the disk, and the condensation
of free-flying PAHs onto the ice mantles
becomes efficient when most of the small
grains are coated with an ice mantle.

%%%%%%%%%%%%%%%%%%%%%%%%%%%%%%%%%%
\section{Discussion}\label{sec:discus}
\subsection{Robustness}\label{sec:robst}

While our standard model reproduces the observed
SED and PAH emission features successfully,
it might not be the unique solution, so we verify
the robustness of our best-fit parameters here.
All the parameters of the porous dust model are
explicitly discussed in Li \& Lunine (2003b)
for the HD\,141569A disk. 
Also, the effects of some parameters 
including those for the dust spatial distribution 
are discussed in Seok \& Li (2015)
for the HD\,34700 disk.
These discussions are expected to be valid 
for the \thisdisk\ disk as well. 

For the PAH modeling, there are three parameters
involved: $a_0$ and $\sigma$ for the size distribution, 
and $\phi_{\rm ion}$ for the ionization fraction. 
With $\phi_{\rm ion}=0.6$, we have tried to fit the observed PAH
emission features with the interstellar PAH size
distribution ($a_0=3.5\Angstrom$, $\sigma=0.4$).
The interstellar mixture also produces a model
spectrum in good agreement with the observed
spectrum. In comparison with our standard model,
as the interstellar mixture contains fewer (more)
small (large) PAHs (see Figure \ref{fig:pahsz}), 
the $3.3\mum$ feature is slightly weaker 
while the $11.3\mum$ feature is slightly stronger. 

For $\phi_{\rm ion}$, the observed spectrum shown
in Figure \ref{fig:sed}
clearly depicts that the PAHs in the \thisdisk~disk
are neither fully ionized ($\phi_{\rm ion}=1.0$)
nor fully neutral ($\phi_{\rm ion}=0$).
Varying $\phi_{\rm ion}$ affects the relative strengths
of the PAH emission bands significantly
(e.g., see Allammandola et al.\ 1999).
We have calculated $\phi_{\rm ion}=0.4$ and 0.8, 
and the $\phi_{\rm ion}=0.4$ model
produces too strong a $3.3\mum$ feature 
by a factor of $\simali$$2$ 
while the $\phi_{\rm ion}=0.8$ model 
emits too weak a $3.3\mum$ feature 
by a factor of $\simali$$2$. 
The relative strength of 
the $11.3\mum$ feature behaves 
in a way similar to that of the $3.3\mum$ feature,
which increases (or decreases) within $\simali$$ 50\%$
for $\phi_{\rm ion}=0.4$ (or $=0.8$).
Note that we assumed a constant $\phi_{\rm ion}$ for
PAHs of all sizes so far, and we will discuss its dependence
on the PAH size in Section \ref{ssbsec:pahchg}.

For the dust modeling, three components (hot, warm, and
cold components) are combined
to explain the observed SED, and each component can be
described by a distinct set of grain size and spatial distributions. 
For all components, our models are not sensitive to the
lower ($\amin$) and upper ($\amax$) cutoff sizes 
of the dust size distribution
except $\amin$ of the warm component
and $\amax$ of the hot component, 
which affect the strength of the silicate emission feature 
at $\simali$$10\mum$ (see Section \ref{sec:dust}).

For the hot component, since the inner ring %($r\la1\AU$)
has not been spatially resolved, its dust spatial distribution
parameters $r_{\rm p, h}$ and $w_{\rm h}$ cannot be firmly
constrained. We vary $r_{\rm p, h}$ between $0.2\AU$
and $0.4\AU$ and $w_{\rm h}$ between $0.1\AU$ and
$0.3\AU$. Any combination within these ranges 
does not alter the model-fit significantly 
and the model emission varies within 5\% 
at $1\mum\la\lambda\la10\mum$.
Models with $r_{\rm p, h}\ga0.4\AU$ show
deficiency in the 2MASS {\it J, H, K$_{\rm s}$}
bands and a stronger $10\mum$ silicate feature.

For the warm component, $r_{\rm p, w}$ and
$w_{\rm w}$ can be directly compared to the middle
ring spatially resolved in the near-IR
polarized light image (\eg Quanz \etal2013), or in
the 7\,mm dust continuum (\eg Osorio \etal2014). 
To be consistent with the resolved ring structure,
we only vary $25\AU\la r_{\rm p, w}\la45\AU$
and $5\AU\la w_{\rm w}\la15\AU$. Models with 
$r_{\rm p, w}\ga40\AU$ slightly overproduce the
dust continuum at wavelengths longer than $25\mum$.
Models with $30\AU\la r_{\rm p, w}\la35\AU$ can also
reproduce the observed SED successfully 
except that as $r_{\rm p, w}$ gets
smaller, the $10\mum$ silicate
feature becomes stronger. 
Varying $w_{\rm w}$ within the above range 
does not alter the model-fit considerably. 
Only for the case with small $r_{\rm p, w}$ 
(\ie $r_{\rm p, w}\sim25\AU$), 
as $w_{\rm w}$ increases, 
the $10\mum$ feature gets even stronger.
This is because more grains are
located close to the central star
(Equation \ref{eq:dndr_w}).
With $r_{\rm p, w}$ and $w_{\rm w}$ fixed,
we examine the power-law index
$\alpha_{\rm w}$ for the size distribution.
It is found that $\alpha_{\rm w}$ mainly affects
the slope of the dust continuum at $15\mum\la\lambda
\la40\mum$. The {\it Spitzer}/IRS spectrum shows that
the dust continuum raises at $15\mum\la\lambda\la
20\mum$ and becomes roughly flat at $\lambda\ga20
\mum$ (see Figure \ref{fig:sed}(c)). Models with flat
$\alpha_{\rm w}$ ($\simali$$3.0$) cannot explain the
flat continuum as there would be too many large grains 
which emit at longer wavelengths. 
Models with $3.3\la\alpha_{\rm w}\la3.7$
give reasonable fits, 
but steeper (shallower) slopes slightly
overproduce the emission at wavelengths shorter
(longer) than $\lambda\sim22\mum$.

Similar to the warm component, 
the spatial distribution of 
the cold component can be directly 
compared to the direct imaging. 
As the middle gap is located at 
$40\AU\la r \la 70\AU$
(\eg Quanz \etal2003, Momose \etal2015),
$r_{\rm p,c}$ is supposed to be near
the outer radius of the gap ($\simali$$70\AU$).
Models with $70\AU\la r_{\rm p,c}\la 80\AU$
give almost identical SEDs except slight
difference at $\lambda\la100\mum$ 
(e.g., less than 10\% of the flux level
at $\lambda\sim60\mum$). We also examine
$\alpha_{\rm c}$ by varying it between 2.5
and 3.5. Models with steeper slopes
($\alpha_{\rm c}\sim3.5$)
emit too much at $\la70\mum$ and too little
at mm wavelengths. Models with shallower
slopes ($\alpha_{\rm c}\sim2.5$) can produce
reasonable fits to the observed SED but emit
more by a factor of $\simali$$2$--4 in the five
JCMT bands ($\lambda\sim450\mum$--$2\mm$). 

Based on previous observations (\eg Grady \etal
2007, Momose \etal2015), $\gamma=1$ is
adopted in our models (see Section \ref{sec:disk}). 
We find that increasing $\gamma$ ($1.0\la\gamma
\la3.0$) does not affect the model-fit significantly
except $M_{\rm dust}$ decreasing by a factor of
$\simali$$3$ at most and a little deficiency at 7\,mm.
In fact, our standard model already shows 
a deficiency at 7\,mm 
by a factor of $\simali$$3$. 
The contribution of free-free emission 
is expected to be negligible
as it is less than 5\% of 
the observed VLA 7\,mm flux density 
(Osorio \etal2014). 
In addition, previous studies (\eg Quanz \etal2013) 
suggest that the radial profiles of 
the observed surface brightness of the outer disk
would be better explained with two power-laws;
the slope of the inner region ($r\la120\AU$) is
relatively shallower than that of the outer region
($r\ga120\AU$). 
These results imply that the outer disk might require
an additional component of the disk geometry and/or
multiple dust grain populations 
to explain the 7\,mm emission.
We prefer not to model the 7\,mm emission 
in terms of a two-power-law dust spatial distribution
as this would inevitably requires three more parameters.

In summary, we have examined various models
with different combinations of parameters and find
that our main results are maintained. Therefore,
the porous dust model together with
the astro-PAH model is robust in explaining the
dust and PAH emission from the \thisdisk~disk.

\subsection{PAHs in the \thisdisk~Disk}\label{subsec:pah}
\subsubsection{Abundance}\label{ssbsec:pahabd}

Because PAHs are exposed to stochastic heating
by starlight (Draine \& Li 2001),
their IR emission profiles are independent of the 
starlight intensity, and the absolute flux levels of the
PAH features simply scale with the starlight intensity. 
Consequently, the total PAH mass, $M_{\rm PAH}$,
is inversely proportional to the starlight intensity ($U$),
%(\ie $M_{\rm PAH}\propto 1/U$), and
and as $U$ is proportional to $r^{-2}$ 
(Section \ref{sec:pah}),
$M_{\rm PAH}$ scales with $r^2$.

To account for the observed fluxes of the PAH features, 
the total PAH mass in the \thisdisk~disk can be expressed as
$M_{\rm PAH}(r)\approx4.55\times10^{-7}(r/\AU)^2\Mearth$
assuming all PAHs are concentrated at a distance of $r$.
Although the spatial distribution of the PAH emission 
along the radial direction of the disk has not been
revealed observationally,  
the spatially extended 3.3 and $11.3\mum$ features
clearly indicate that it is extended to up to
$\simali$$50\AU$ (Habart \etal2006, Maaskant \etal2013).
Adopting $r=50\AU$ results in
$M_{\rm PAH}\approx1.14\times10^{-3}\Mearth$
(or $3.42\times10^{-9}\Msun$),%\footnote{%
which is about a factor of $\simali$$10$
smaller than the previous estimates
(\eg $M_{\rm PAH}\sim4.0\times10^{-8}\Msun$
in Maaskant \etal2014 or $\simali$$7.5\times10^{-8}\Msun$
in Meeus \etal2010). 
Since a large amount of PAH molecules could
have frozen onto the ice mantles of cold dust grains 
beyond $\simali$$50\AU$ (\ie snowline),
our estimate of $M_{\rm PAH}$ may not
represent the actual total PAH mass in this disk. 
This could be the primary reason for the difference between our
estimate and those of the previous studies, 
which assumed a constant PAH-to-dust
mass fraction over the entire outer part of the disk
(including the middle ring and the outer disk, $r\ga20\AU$). 
The spatial distribution of PAHs in the \thisdisk~disk is
critical to calculate $M_{\rm PAH}$,
which still remains
unknown in detail except the outer boundary
of the PAH emission.
Future near- and mid-IR observations 
with higher spatial resolution
and higher sensitivity could allow us 
to place a stronger constraint on 
the current PAH modeling.

The deficit of PAHs in protoplanetary disks or
debris disks has been reported in the literature
(\eg Li \& Lunine 2003b, Geers \etal2006,
Thi \etal2013, Seok \& Li 2015). Indeed,
a low PAH mass fraction of the \thisdisk~disk
has also been found by previous studies
(\eg $M_{\rm PAH}/M_{\rm dust}\sim5\times10^{-4}$, 
Meeus \etal2010, Maaskant \etal2014).
%Recalling $M_{\rm dust}\approx176\Mearth$
As our estimate of $M_{\rm PAH}$ does not
constrain the PAH mass beyond the snowline
(\ie $r\approx50\AU$), we compare $M_{\rm PAH}$
with the total mass of the hot and warm components
(\ie $\approx9.15\times10^{-1}\Mearth$, see Section \ref{sec:result}),
which yields a PAH-to-dust mass ratio of
$M_{\rm PAH}/M_{\rm dust}\approx1.25\times10^{-3}$.
%$M_{\rm PAH}/M_{\rm dust}\approx6.47\times10^{-6}$.
Although our ratio is higher than the previous estimates
by a factor of $\simali2$, 
it still indicates that the \thisdisk~disk 
has a deficit of PAHs relative to dust
compared to the typical PAH abundance in the ISM
($M_{\rm PAH}/M_{\rm dust}\approx0.05$,
Li \& Draine 2001, Draine \& Li 2007).
The origin of the deficiency of PAHs remains unclear.
Performing a systematic
study of the PAH emission in protoplanetary disks and
debris disks would extend our understanding of
the origin of PAHs and their characteristics in the disks
(J. Y. Seok \& A. Li 2015, in preparation).

%%%%%%%%%%%%%%%%%%%%%%%%%%%%%%%%%

\subsubsection{Photodestruction}\label{sec:phot}

\begin{figure}[tbp]
\epsscale{0.50}
\plotone{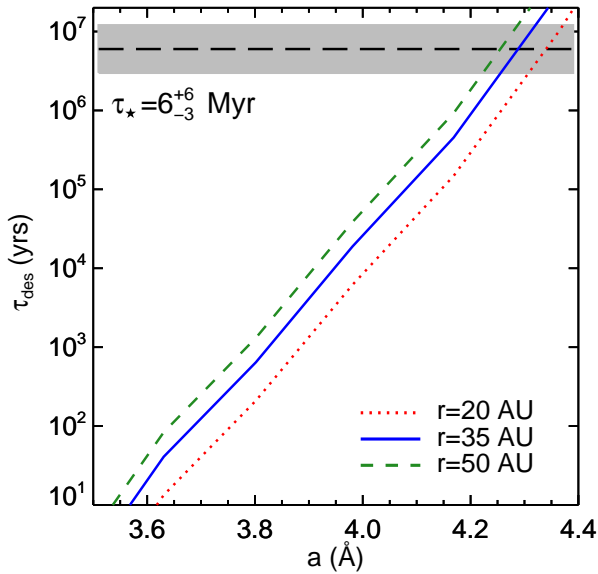}
\caption{\footnotesize
         \label{fig:kdes}
         Photodestruction timescales 
         ($\tau_{\rm des}$) for PAHs 
         at $r=20\AU$ (dotted line),
         $35\AU$ (solid line), and
         $50\AU$ (dashed line)
         as a function of size. 
         The age of \thisdisk~with its uncertainties
         ($\tau_\star=6^{+6}_{-3}\Myr$) 
         is designated by a long-dashed line 
         with a shaded area. 
         }
\end{figure}

When an energetic photon hits a PAH molecule,
this could eventually result in its photodissociation.
In particular for small PAHs that do not have enough 
internal vibrational modes to distribute the energy 
absorbed from the photon, 
they may eject a hydrogen atom,
a hydrogen molecule,
or an acetylene molecule (C$_2$H$_2$)
and end up with dissociation. 
We refer to Li \& Lunine (2003b, see their Appendix A) 
for a detailed discussion of the PAH photophysics.
Following their approach, we assume that a single 
ejection of an acetylene molecule causes
the PAH destruction and derive
the photodestruction rates ($k_{\rm des}$)
of PAHs exposed to UV photons in the \thisdisk~disk.
Figure \ref{fig:kdes} represents the timescales of
photodestruction ($\tau_{\rm des}\equiv1/k_{\rm des}$)
for small PAHs at specific
distances from the central star
($r=20$, 35, and $50 \AU$). 
PAHs smaller than $\simali$$4.3\Angstrom$
(or those containing the number of carbon atoms 
$N_{\rm C}$ fewer than $\simali$$37$) are
likely to be destroyed during the lifetime
of \thisdisk~($\tau_\star=6$ Myr).
This indicates that a continuous replenishment
of small PAHs is required to explain the
current abundance, which is evidenced by
the observed $3.3\mum$ PAH feature.

To maintain the population of small PAHs
in the region where the $3.3\mum$ PAH feature
has been detected (including
the inner gap and the middle ring),
continuous replenishment has been occurring
at a rate of
\beq\label{eq:pahdes}
\dot{M}_{\rm PAH}=\int^{50 \AU}_{\rmin}\sigma_{\rm PAH}(r)
2\pi r~dr 
\times\int^{\infty}_{\apahmin}da
\frac{dn_{\rm PAH}}{da}\frac{(4\pi/3)a^3\rho_{\rm PAH}}{\tau_{\rm des}(a, r)},
\eeq
where $\apahmin$ is the minimum size of PAHs,
and $\rho_{\rm PAH}$ is the mass density of PAHs,
which is adopted to be that of graphite
($\approx2.24 \g\cm^{-3}$).
The PAH surface density distribution,
$\sigma_{\rm PAH}(r)$, is assumed to be the same
as the spatial distribution of the warm dust component
with a cutoff at $r=50\AU$
(Equation \ref{eq:dndr_w}, see Section \ref{sec:disk}):
\beq\label{eq:pahdndr}
\sigma_{\rm PAH}(r)=\sigma^{p}_{\rm PAH} \exp
\left[-4 \ln2 \left(\frac{r-r_{\rm p, w}}{w_{\rm w}} \right)^2 \right] ,
\eeq
where $\sigma^{p}_{\rm PAH}$ is the PAH mid-plane
surface density at $r_{\rm p, w}$.

Adopting $a_0=3.0\Angstrom$, $\sigma=0.4$,
$\apahmin=3.5\Angstrom$, $\rmin=0.2\AU$,
$r_{\rm p, w}=35\AU$, and $w_{\rm w}=10\AU$
from the standard model, we calculate 
$\sigma^{p}_{\rm PAH}\approx 5.46\times10^{22}\cm^{-2}$,
which yields the PAH replenishment rate of
$\dot{M}_{\rm PAH}\approx2.37\times10^{-6} \Mearth\yr^{-1}$
taking only photodestruction into account.

%%%%%%%%%%%%%%% Figure 7: PAH destruction %%%%%%%%%%%%%%%
%%%%%%%%%%%%%%% Figure 7: PAH destruction %%%%%%%%%%%%%%%

In addition to photodissociation,
PAHs as well as dust grains in the \thisdisk~disk are subject to
radiation pressure, Poynting-Robertson drag, and gas drag.
While the stellar radiation pressure which blows away materials
is effective for dust with a size of $a\la10\mum$,
the Poynting-Robertson drag is effective for dust with a size
of $10\mum\la a \la20\mum$ and causes the dust to spiral
inward to the star. As this disk is still gas-rich, the gas drag
dominates the motion of dust grains. Our simple estimate of
a gas-drag timescale is very short (only a few decades),
and a full treatment of the
gas-drag requires a 2-dimensional disk structure
(\eg see Chiang \& Yudin 2010 for a review), which is beyond
the scope of this paper.

\subsubsection{Ionization}\label{ssbsec:pahchg}

%%%%%%%%%% Figure 8: PAH models with phi(a) %%%%%%%%%%%
\begin{figure*}[htbp]
\epsscale{1.00}
\plotone{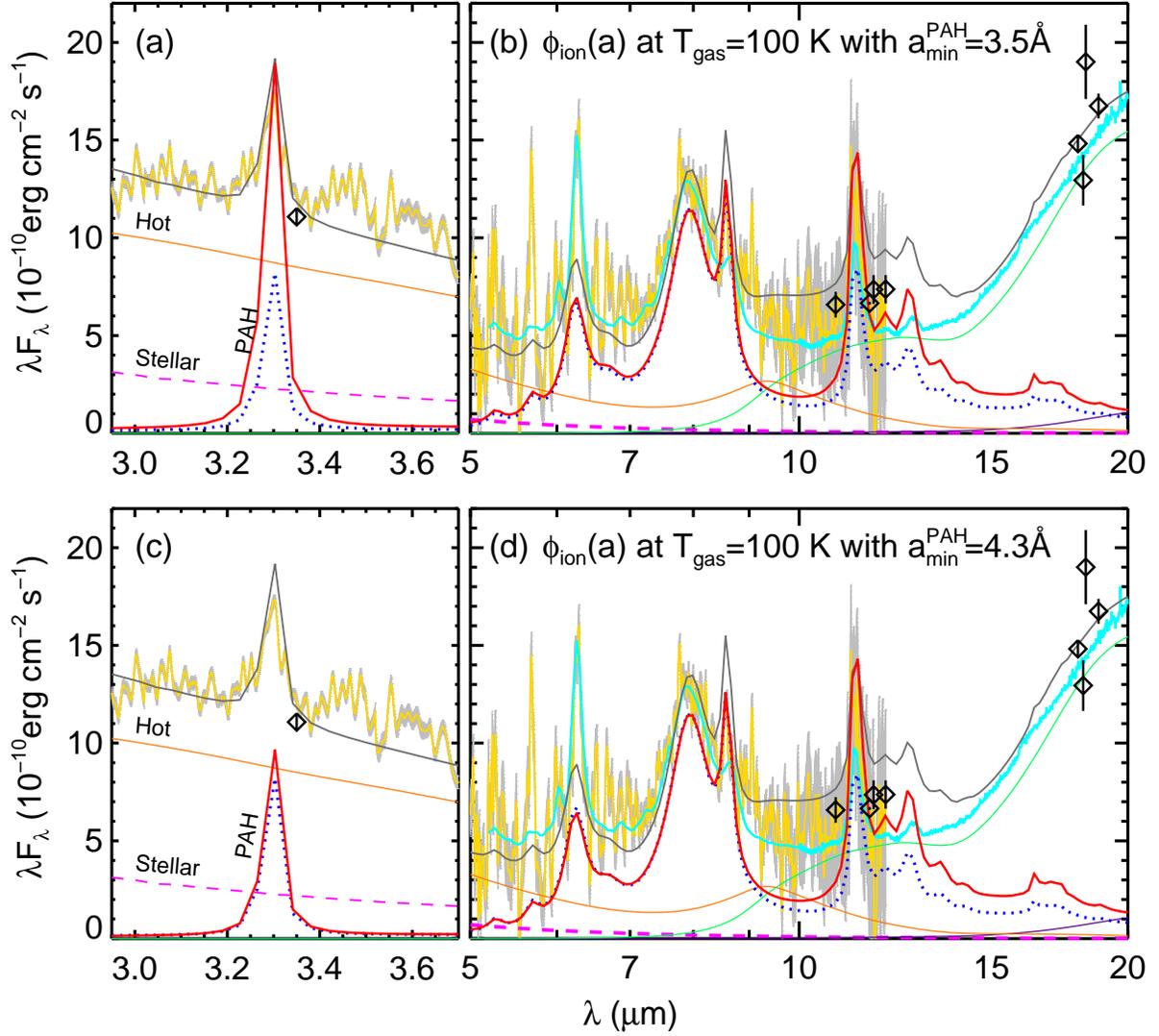}
\caption{\footnotesize
         \label{fig:pahchg}
         Same as Figure \ref{fig:sed}(b) and (c), except
%        best fits to the {\it ISO}/SWS and {\it Spitzer}/IRS
%        spectra provided by 
         that the ionization fraction of PAHs,
         $\phi_{\rm ion}$, as a function of PAH size ($a$)
         is taken into account.
         $T_{\rm gas}=100 \K$ is assumed, which
         corresponds to a distance of $\simali$$24\AU$
         (see Figure \ref{fig:ne}). The PAH size distribution
         used in the standard model ($a_0=3.0\Angstrom$
         and $\sigma=0.4$) gives the best fit.
         For comparison, 
         the standard model with the constant ionization
         fraction (\ie $\phi_{\rm ion}=0.6$) is overlaid.
         Color schemes are
         identical to those in Figure \ref{fig:sed} except that
         the total spectrum of the standard model
         is shown with a dark grey line and the best-fit
         PAH models with variable $\phi_{\rm ion}$ are 
         denoted with a red solid line. Panels (a) and (b)
         are for the PAH component with
         the same lower cutoff size as the standard model
         (\ie $\apahmin=3.5\Angstrom$), while panels (c)
         and (d) are for a larger lower cutoff size
         of $\apahmin=4.3\Angstrom$. In comparison
         with the standard model,
         the $\apahmin=3.5\Angstrom$ model predicts
         a much stronger $3.3\mum$ feature, whereas
         the $\apahmin=4.3\Angstrom$ model is very
         similar to the standard model, which reproduces
         the observed 3.3$\mum$ feature very well. 
         Both models with variable $\phi_{\rm ion}$
         overproduce the $11.3\mum$ feature. 
         }
\end{figure*}
%%%%%%%%%% Figure 8: PAH models with phi(a) %%%%%%%%%%%

The charge state of PAHs is
determined by the balance between the photoelectric
emission and collisions with electrons and ions. 
The steady state charge distribution as a function
of PAH size ($a$) can be calculated
from the photoemission rate, the positive ion
accretion rate, and the electron accretion rate
when relevant ambient conditions are given
(\eg see Bakes \& Tielens 1994,
Weingartner \& Draine 2001). 
Since the detailed information 
about the electron density
($n_e$) such as its spatial distribution
is not well known for the \thisdisk~disk,
we adopt a constant ionization fraction for our
standard model, and $\phi_{\rm ion}=0.6$
gives the excellent fit to the observed PAH
emission features. Here, we explore the effect of
the ionization fraction more by calculating
$\phi_{\rm ion}$ as a function of PAH size ($a$)
and adopting the radial profiles of $n_{\rm e}$ and
$U$ shown in Figure \ref{fig:ne}. 
As the absorption cross sections for PAH molecules
are proportional to $a^3$
(\ie $C^{\rm PAH}_{\rm abs}\propto a^3$),
it is naturally expected
that large PAHs are more ionized while small ones
are more neutral. This is also consistent with
the comparison between the photoionization
and recombination timescales (Figure \ref{fig:ion}).

Figure \ref{fig:pahchg} shows the best-fits
to the {\it ISO}/SWS and {\it Spitzer}/IRS spectra
provided by the PAH models with the ionization 
fraction $\phi_{\rm ion}(a)$ calculated from
the balance between the photoionization and 
electron recombination. We adopt
a gas temperature $T_{\rm gas}=100\K$, which
corresponds to that at $\simali$$24\AU$ 
(see Figure \ref{fig:ne}). 
Within the range of radial distances where PAHs
are possibly distributed in the \thisdisk~disk
(\ie a few $\AU\la r \la 50\AU$), $T_{\rm gas}$
varies by less than a factor of 2, and
the resultant PAH spectra are not sensitive to
$T_{\rm gas}$ in the context that the ionization
parameter is only proportional to $T^{1/2}_{\rm gas}$.
The best-fit shown in Figure \ref{fig:pahchg} is obtained
with the PAH size distribution used in the standard
model ($a_0=3.0\Angstrom$ and $\sigma=0.4$).
All other parameters associated with dust modeling 
are identical to those in the standard model.

In comparison with the standard model 
for which the PAH emission is shown 
as a blue dotted line in Figure \ref{fig:pahchg}, 
the best-fit model with variable $\phi_{\rm ion}$ 
increases the strength of 
the 3.3 and $11.3\mum$ features 
while the 6.2, 7.7, and $8.6\mum$ features 
are not altered. 
This is because small PAHs that dominate
the the 3.3$\mum$ emission feature become more
neutral (\ie $\phi_{\rm ion}(a)\sim0.1$--0.3 for
$a\la5\Angstrom$) than those in the standard model
with $\phi_{\rm ion}=0.6$; moreover, neutral PAHs
also emit more strongly at the 11.3$\mum$ feature
than ionized PAHs. 
To avoid the overproduction of these features, 
we adopt the lower cutoff size
($\apahmin$) for the PAH component from the largest
PAH size that can be destroyed by photodissociation
during the lifetime of \thisdisk~%($\simali$$6\Myr$)
(\ie $\apahmin=4.3\Angstrom$, see Section
\ref{sec:phot} and Figure \ref{fig:kdes}).
The best-fit model with $\apahmin=4.3\Angstrom$
predicts the strength of the $3.3\mum$ feature 
to be comparable to that of the standard model, 
which successfully reproduces the observed feature
(see Figure \ref{fig:pahchg}(c)). However, the intensity
of the $11.3\mum$ feature still remains overestimated.
This is attributed to the overpopulation of neutral
PAHs with a size of $4.3\Angstrom \la a \la 10\Angstrom$,
which is the size regime of neutral PAHs that efficiently 
radiate the $11.3\mum$ feature (\eg Draine \& Li 2007).

We do not attempt any further adjustment to reproduce
the observed $11.3\mum$ feature since $\phi_{\rm ion}(a)$
is currently calculated on the basis of uncertain
ambient conditions (\eg $n_{\rm e}$). It is noteworthy
that our standard model with $\phi_{\rm ion}=0.6$
gives an excellent fit to the observed PAH emission,
which might indicate that more small PAHs are
positively or negatively charged in the \thisdisk~disk. 
We speculate that anionic small PAHs are more
favorable because small PAHs are required to
be close to the central star to be ionized
(see Figure \ref{fig:ion}) where they are subject to
more efficient photodissociation (see Figure \ref{fig:kdes}).
Further knowledge on the ambient conditions and
the spatial variation of the relative strengths 
the of PAH features can verify the true physical 
properties of PAHs in the \thisdisk~disk.

\subsubsection{Aliphatics versus Aromatics}

In addition to the major, prominent PAH features
at 3.3, 6.2, 7.7, 8.6, 11.3, and 12.7$\mum$,
several minor spectral features at 3.43, 6.87, 
and 7.23$\mum$ are also clearly seen in 
the {\it ISO}/SWS and {\it Spitzer}/IRS spectra 
of the \thisdisk~disk 
(see Figure \ref{fig:sed}b,c;
also see Sloan \etal2005, Acke \etal2010).
These features are generally attributed to 
the C--H vibrational modes in aliphatic hydrocarbons 
(\eg Chiar \etal2000).
The detection of these aliphatic features 
indicates that the PAH species in the \thisdisk~disk 
are alkylated and have aliphatic CH sidegroups.

Following Li \& Draine (2012), 
we estimate the aliphatic fraction of 
the PAH molecules in the \thisdisk~disk. 
Let $I_{3.3}$ and $I_{3.4}$ respectively be 
the total (integrated) fluxes emitted from
the 3.3 and $3.43\mum$ features.
The {\it ISO}/SWS spectrum gives
$I_{3.4}/I_{3.3}\approx0.48$. 
Assuming that the $3.43\mum$ feature 
exclusively originates from 
the aliphatic C--H bond 
(\ie no contribution from
anharmonicity (Barker \etal1987)
or superhydrogenation
(Bernstein \etal1996)),
we derive an upper limit on the ratio
of the number of C atoms in aliphatic sidegroups
to that in aromatic benzene rings from
$N_{\rm C,\,aliph}/N_{\rm C,\,arom}\approx
0.3\times(I_{3.4}/I_{3.3})\times(A_{3.3}/A_{3.4})$,
where $A_{3.3}$ and $A_{3.4}$
are the band strengths of the aliphatic and
aromatic C--H bonds, respectively.
The coefficient of 0.3 comes from 
the assumption that one aliphatic C atom 
corresponds to 2.5 aliphatic C--H bonds 
while one aromatic C atom corresponds to 
0.75 aromatic C--H bond (see Li \& Draine 2012).
With $A_{3.4}/A_{3.3}\approx1.76$
(Yang \etal2013), we obtain
$N_{\rm C,\,aliph}/N_{\rm C,\,arom}\approx0.082$.
This implies that the PAH molecules 
in the \thisdisk~disk are mostly aromatic.

Similarly, let $I_{6.9}$ and $I_{7.7}$ respectively
be the measured intensities 
of the 6.87 and $7.7\mum$ emission features 
as shown in the {\it Spitzer}/IRS spectrum.
By fitting the features with Drude
profiles, we obtain $I_{6.9}/I_{7.7}\approx0.01$.
%which is also consistent with what we calculate with
%the flux measurement of these features listed in
%Acke \etal(2010). 
Let $A_{6.9}$ and $A_{7.7}$ respectively 
be the band strengths 
of the aliphatic C--H deformation band 
and the aromatic C--C stretching band.
Let $B_{\lambda}(T)$ be the Planck function 
at wavelength $\lambda$ and temperature $T$.
For $330\le T\le 1000\K$ we obtain
$B_{6.9}/B_{7.7}\approx0.9\pm0.2$.
With $A_{6.9}\approx2.3\times10^{-18}\cm$ 
per CH$_2$ or CH$_3$ group and
$A_{7.7}=5.4\times10^{-18}\cm$ 
per C atom for charged aromatic molecules
(see Li \& Draine 2012),
we derive the ratio of
the number of carbon atoms in aliphatic sidegroups
to that in aromatic benzene rings:
$N_{\rm C,\,aliph}/N_{\rm C,\,arom}\approx
(I_{6.9}/I_{7.7})\times(A_{7.7}/A_{6.9})\times
(B_{7.7}/B_{6.9})\approx0.025$.
Again, this demonstrates that the PAH molecules 
in the \thisdisk~disk are predominantly aromatic.

\subsubsection{Links to Planet Formation}

As recent observations have revealed
point-like features in the \thisdisk~disk
(\eg Biller \etal2014, Osorio \etal2014,
Reggiani \etal2014) together with
the nature of pre-transitional disks
such as a cavity with an inner disk,
we would postulate that planet formation
is ongoing in the \thisdisk~disk so
that we can briefly discuss the link 
between PAH molecules and planet formation. 

PAHs play an important role in the physical
and chemical evolution of dust disks
around young stellar objects, where planet
formation takes place. PAH molecules
in the surface layer of a dust disk where
the stellar UV radiation directly reaches absorb UV
photons efficiently and get photoionized.
Energetic electrons ejected from the ionized PAHs
are the dominant heating source of the
surrounding gas, of which the temperature
in turn determines the vertical structure
of the disk. The energy absorbed by PAHs
reradiate in their IR bands, which influence 
the physical processes driven
by radiation close to the midplane.
Chemically, both free-flying PAHs as well
as those frozen-out on ice-mantles of
dust grains are important. Neutral and
negatively charged PAHs can transfer
electrons with C$^+$ efficiently, which
affect the carbon chemistry of the disk.
Ice mantles where PAHs are frozen can be
processed by UV light and cosmic-rays
so that various complex (organic) molecules
can form (\eg see Bernstein \etal2002,
Bouwman \etal2011).
Since such ice-coated grains are
the building blocks of comets and planetesimals,
these materials could be delivered to newly-forming
planets and affect their early phase of the
chemical evolution. Although it is observationally
established that the 3.3 and 11.3$\mum$ PAH emission
is spatially extended in the \thisdisk~disk, 
the precise spatial distribution of PAHs
has not been resolved, so we cannot directly 
connect PAHs to the ongoing planet formation. 
However, the \thisdisk~disk provides an unique 
opportunity to probe the role of PAHs 
in planet formation,
and high spatial resolution studies by
the Atacama Large Millimeter/Submillimeter Array
(ALMA) and the {\it James Webb Space Telescope} ($JWST$)
will allow us to unveil the mutual interaction of
the gas, dust, and PAHs in the disk and
their roles in the ongoing planet formation.

\section{Summary}\label{sec:sum}

We have modeled the PAH emission features and
the dust thermal emission SED 
from the near-IR to submm/mm wavelengths 
of the pre-transitional disk around
\thisdisk~with porous dust aggregates 
and a mixture of neutral and charged PAHs. 
Taking into account the spatially resolved 
disk structure (inner cavity, middle ring, 
middle gap, and outer disk) together with 
an inner ring inferred from 
the $\simali$2--6$\mum$ near-IR excess emission, 
our standard model,
consisting of three dust components
and relatively small PAH molecules
(following a log-normal size distribution with
$a_0=3.0\Angstrom$ and $\sigma=0.4$),
%compared with those in the diffuse ISM,
provides an excellent fit to
the entire SED and the PAH emission features. 
%[most important finding?]

The observed {\it Spitzer}/IRS and {\it ISO}/SWS 
IR spectra of the \thisdisk~disk show 
the prominent PAH band emission
at 3.3, 6.2, 7.7, 8.7, 11.3, and $12.7\mum$, 
which can be reproduced by PAHs 
with an ionization fraction of
$\phi_{\rm ion}\sim0.6$ for all PAH sizes. 
A comparison between the
photoionization and electron recombination timescales
suggests that such a high $\phi_{\rm ion}$ can be 
attributed to either positively charged PAHs located
at the inner gap ($r\la15\AU$) or negatively
charged PAHs in the middle ring ($25\AU\la r \la 40\AU$)
where the electron density is high. 
As the 3.3 and $11.3\mum$
PAH features are known to extend up to
$\simali$$50\AU$ in the disk, both cases are plausible.
For comparison with the constant $\phi_{\rm ion}$
scenario, $\phi_{\rm ion}(a)$ as a function of
PAH size ($a$) at a given distance from the central star
is calculated according to the ambient physical conditions.
As small PAHs tend to be more neutral than large PAHs,
models with variable $\phi_{\rm ion}(a)$ overproduce 
the intensities of the 3.3 and $11.3\mum$ features 
when the lower cutoff of the PAH size 
is the same as the standard model 
(\ie $\apahmin=3.5\Angstrom$). 
Adopting the largest PAH size 
subject to the photodestruction
during the lifetime of \thisdisk~as the lower cutoff
size (\ie $\apahmin=4.3\Angstrom$), the observed
$3.3\mum$ feature is reproduced closely while
the $11.3\mum$ feature is still enhanced.
This implies that more small PAHs are required
to be charged either positively or negatively
in the \thisdisk~disk to explain the observed
$11.3\mum$ emission.
 
The minor PAH features at 3.43, 6.87, and $7.23\mum$
are also clearly detected in the {\it ISO}/SWS
and {\it Spitzer}/IRS spectra.
These minor features are indicative of 
aliphatic sidegroups attached to 
the benzene rings of PAHs.
The 3.43$\mum$ feature is generally attributed
to the aliphatic C--H stretching band, 
while the 6.87 and 7.23$\mum$ features are
attributed to the aliphatic C--H deformation bands.
By comparing the intensity of the 3.43$\mum$ feature
to that of the 3.3$\mum$ feature, 
we place an upper limit of $\simali$$0.082$ 
on the aliphatic fraction of the PAH molecules
in the \thisdisk~disk. 
Similarly, an aliphatic fraction of $\simali$0.025
is derived by comparing the intensity of 
the 6.87$\mum$ feature to that of the 7.7$\mum$ feature. 
We conclude that the PAH molecules in the \thisdisk~disk 
are predominantly aromatic.
%aliphatic fraction -- ratio of 
%C atoms in aliphatic sidegroups 
%to that in aromatic benzene rings

The three dust components 
in the standard model include
a hot dust component 
($\alpha_{\rm h}=3.5$, $0.01\mum\la a \la 1\mum$),
a warm dust component 
($\alpha_{\rm w}=3.5$, $1\mum\la a \la 1\cm$), 
and a cold dust component 
($\alpha_{\rm c}=3.0$, $1\mum\la a \la 1\cm$),
which mainly originate from the inner ring, the
middle ring, and the outer disk, respectively.
Our standard model suggests that the dust
distribution in the inner disk peaks at
at $r_{\rm p, h}=0.3\AU$ with a FWHM of
$w_{\rm h}=0.1\AU$.
The spatial distributions of the warm and cold
components ($r_{\rm p, w}=35\AU$ with
$w_{\rm w}=10\AU$, and $r_{\rm p, c}=80\AU$)
are in good agreement with
the disk geometry previously resolved by direct
imaging observations. The total dust mass
in the disk is calculated to be $\approx176\Mearth$,
which is dominated by the cold dust component ($\simali$99.48\%).

The total IR luminosity of $L_{\rm IR}\approx3.5\Lsun$
is obtained for a distance of $d=145\pc$, of which
$\simali$10\% is contributed by the PAH emission.
The hot, warm, and cold components 
account for $\simali$31\%, 40\%, and 19\% 
of the total IR luminosity, respectively.

In thermal equilibrium, ice mantles of porous dust
grains in the \thisdisk~disk are sublimated
between $\simali$$16$ and $60\AU$.
Beyond $\simali$$40\AU$, grains even smaller than $5\mum$
become ice-coated, so that free-flying PAH
molecules can easily stick onto the ice mantles of
the grains. This is consistent with the previous
observations showing that
the 3.3 and $11.3\mum$ emission features
are extended up to $\simali$$50\AU$, which implies
that the snowline is closely related to the spatial
distribution of the observable PAH emission. 
Assuming that the all the PAHs are located at
a distance of $r$ from the central star,
the observed PAH emission yields a
total PAH mass of $M_{\rm PAH}\approx4.55
\times10^{-7}(r/\AU)^2\Mearth$.
Adopting $r=50\AU$, $M_{\rm PAH}\approx
1.14\times10^{-3}\Mearth$ is obtained,
which yields a PAH-to-dust mass ratio of
$M_{\rm PAH}/M_{\rm dust}\approx
1.26\times10^{-3}$ in comparison with the total
mass of the hot and warm components.
This indicates a deficit of PAHs
(relative to dust) in the \thisdisk~disk with
respect to the ISM,
which is a common case for dust disks 
around young stars. 

PAH molecules, in particular small PAHs,
are subject to photodissociation by absorbing
an energetic photon. During the lifetime of
the \thisdisk~disk ($\simali$$6\Myr$), PAHs
smaller than $\simali$$4.3\Angstrom$ are photo-destroyed,
and to maintain the population of small PAHs,
a PAH mass replenishment rate of
$\dot{M}_{\rm PAH}\approx2.37\times10^{-6}\Mearth
\yr^{-1}$ is required.
By combining the distinct PAH emission features
and the spatially resolved disk structure with
the detection of point-like sources in the disk,
\thisdisk~offers a unique opportunity to 
study the role of PAHs in disk physics and
ongoing planet formation.  

\acknowledgements
We thank M. J. Barlow, C. M. Telesco,
P. Woitke, and H. Zhang for very helpful discussions.
We are grateful to the anonymous referee for
his/her valuable comments that improved our manuscript.
We are supported in part by
NSF AST-1109039, 
NNX13AE63G, 
%NSFC\,11173007, 
NSFC\,11173019, 
%NSFC\,11273022,
and the University of Missouri Research Board.

%%%%%%%%%%%%%%% References %%%%%%%%%%%%%%%%%%%%%%%%%%%%
%

\end{document}